\documentclass{comnet}

\usepackage{amsmath} 
\usepackage[caption=false]{subfig} 
\usepackage{xcolor}

\usepackage{graphicx}
\usepackage{dcolumn}
\usepackage{bm}
\usepackage{siunitx}

\usepackage{array}
\newcolumntype{L}[1]{>{\raggedright\let\newline\\\arraybackslash\hspace{0pt}}m{#1}}
\newcolumntype{C}[1]{>{\centering\let\newline\\\arraybackslash\hspace{0pt}}m{#1}}
\newcolumntype{R}[1]{>{\raggedleft\let\newline\\\arraybackslash\hspace{0pt}}m{#1}}

\begin{document}

\title{The role of bipartite structure in R\&D collaboration networks}

\shorttitle{Bipartite collaboration networks} 
\shortauthorlist{D. Vasques Filho and D.R.J. O'Neale} 

\author{
\name{D. Vasques Filho$^{\dag}$}
\address{Te P\={u}naha Matatini, Department of Physics, University of Auckland, Private Bag 92019, Auckland, New Zealand}
\address{Leibniz-Institut f\"ur Europ\"aische Geschichte, Alte Universit\"atsstra{\ss}e 19, 55116 Mainz, Germany}
\email{$^\dag$Corresponding author: vasquesfilho@ieg-mainz.de}
\and
\name{Dion R.J. O'Neale}
\address{Te P\={u}naha Matatini, Department of Physics, University of Auckland, Private Bag 92019, Auckland, New Zealand}
\email{d.oneale@auckland.ac.nz}
}

\maketitle

\begin{abstract}
{A great number of real-world networks are, in fact, one-mode projections of bipartite networks comprised of two different types of nodes. In the case of interactions between institutions engaging in collaboration for technological innovation, the underlying network is bipartite with institutions (agents) linked to the patents they have filed (artifacts), while the projection is the co-patenting network.
Since projected network properties are highly affected by the underlying bipartite structure a lack of understanding of the bipartite network has consequences for the information that might be drawn from the one-mode co-patenting network.
Here, we create an empirical bipartite network using data from 2.7 million patents recorded by the European Patent Office. We project this network onto the agents (institutions) and look at properties of both the bipartite and projected networks that may play a role in knowledge sharing and collaboration. We compare these empirical properties to those of synthetic bipartite networks and their projections.
We show that understanding the bipartite network topology is critical for understanding the potential flow of technological knowledge. Properties of the bipartite structure, such as degree distributions and small cycles, affect the topology of the one-mode projected network --- specifically degree and clustering distributions, and degree assortativity. We propose new network-based metrics as a way to quantify how collaborative agents are in the collaboration network. We find that several large corporations are the most collaborative agents in the network, however such organizations tend to have a low diversity of collaborators. In contrast, the most prolific institutions tend to collaborate relatively little but with a diverse set of collaborators. This indicates that they concentrate the knowledge of their core technical research while seeking specific complementary knowledge via collaboration with smaller institutions.
}

{Collaboration; patents; bipartite networks; projected networks; co-patenting; network structure}
\end{abstract}

\section{\label{sec:introduction}Introduction}

Many systems represented by networks are one-mode projections of more complicated structures \cite{souma2003complex,newman2001scientific,onody2004complex,watts1998collective}. Often, the original network has a bipartite architecture comprised of two different types of nodes. The topology of the projected network and the dynamics that take place on it can be highly dependent on the features of the original bipartite structure. 
One important example of such bipartite structures are collaboration networks. These include co-authorship networks where researchers jointly publish scientific articles together \cite{acedo2006co,abbasi2011identifying,mali2012dynamic,wagner2005network,glanzel2004analysing,liu2005co} or where institutions may jointly file patent applications (co-patenting networks \cite{schilling2007interfirm,singh2005collaborative,wang2014knowledge,zheng2014international}). 

The body of studies on inter-firm relationships with respect to research and development (R\&D) has increased substantially in the last decades along with the development of network theory \cite{ozman2009inter,schilling2007interfirm}, as ``it is now accepted that innovation is most effectively undertaken as a collective process in which networks play a central role'' \cite{ozman2009inter}. There are several underlying reasons for which institutions might make alliances aiming to increase their capacity to innovate and to create new technologies. Among these reasons we can cite the increasing complexity level and multidisciplinarity of new technologies \cite{hagedoorn1993understanding,arora1994evaluating,rothaermel2004exploration}, reduction of risk and costs \cite{hagedoorn1993understanding,eisenhardt1996resource,wernerfelt1984resource,gulati1998architecture}, reduced time between innovation and market introduction \cite{hagedoorn1993understanding,eisenhardt1996resource,rothaermel2004exploration}, internationalization \cite{hagedoorn1993understanding}, and market vulnerability \cite{hagedoorn1993understanding,eisenhardt1996resource,rothaermel2004exploration}.

Research on inter-firm networks can have different foci as, for instance, how the position of an institution in the network affects its performance, profitability and other strategic management measures \cite{uzzi2002knowledge,shipilov2005should,uzzi1996sources}, network formation based on geographic proximity and its implications \cite{gulati1999network,huggins2010knowledge}, knowledge diffusion \cite{gulati1999network,huggins2010knowledge,huggins2010forms,sammarra2008heterogeneity,grant2004knowledge}, and others. Here, our interest lies in the general structure of collaboration networks and in what the topology of such a structure can tell us about the strategies with which institutions cooperate with each other.

A good example of inter-firm collaboration is the joint application for new patents by multiple institutions (e.g. corporations, universities, government agencies and so on). In such cases, it is possible to construct a bipartite network in which one set of nodes is institutions (agents) that are connected to patents they have applied for (artifacts). We create an empirical bipartite network using data from the European Patent Office dating back to 1978 for 40 countries with harmonized applicant names (OECD, HAN database, February 2016) \cite{patstat}. We project this network onto the agents --- institutions developing patents --- to create a new one-mode network connecting institutions that have patented together. 

We look at the properties of the empirical network that may play a role in knowledge sharing, such as cycles and clustering. It has been shown that clustering ``enables even a globally sparse network to achieve high information transmission capacity through locally dense pockets of closely connected firms'' \cite{schilling2007interfirm}. In order to better understand the role played by bipartite network properties such as degree distribution and degree assortativity in real-world collaboration networks,  we compare empirical network properties to the properties of synthetic bipartite networks and their projections. Synthetic networks are important tools in understanding the processes that might operate in real-world networks since they afford us the opportunity to study multiple versions of a network with specific similar properties. 

To do this, we create synthetic analogues of the empirical network by using a configuration model. This preserves the original degree sequence of both sets of nodes and rewires the links randomly. We also create synthetic networks using a random model \cite{erdds1959random,erdos1960evolution,gilbert1959random}, keeping only the original total degree of each set of nodes. We then compare the structural properties of the empirical network with those of the synthetic bipartite networks and their projections. With both empirical and synthetic networks in hand, it is possible to see the role of the network features in shaping the topology of the projected collaboration network. Small cycles, namely four- and six-cycles, change the expected degree and clustering distributions, respectively. The first by increasing the link weight of nodes that are repeatedly collaborating with the same collaborators. The second by increasing clustering due to transitivity.

We propose new global and local metrics using network features from the bipartite and projected collaboration network. Based on the degree of an agent and the degree of the artifacts each agent is connected to, it is possible to quantify the \textit{collaborativeness} between nodes in the network. Along with this, we introduce an easy way to calculate the \textit{diversity} of a node relative to its collaborators (i.e. whether a particular institution tends to share patents with many different or repeatedly with a small set of other institutions), by using node degree and node strength values in projected networks. 

With these metrics, we show that the most collaborative institutions in the co-patenting network are usually different international branches of the same corporation, co-filing patent applications. These same institutions typically have a low diversity of collaborators or, in other words, they are always patenting among themselves. Such behavior raises the question of whether this is just a ``head office effect'' or indeed a collaboration with a flow of knowledge between the entities involved. On the other hand, the institutions with the largest number of patents in the database (high productivity) have significantly higher diversity of collaborators than those with the highest collaborativeness. That is, although they are relatively less collaborative, the collaboration takes place with a diverse set of collaborators. This suggests that such institutions are seeking to access specific technical knowledge owned by smaller external institutions.

The remainder of this paper will be organized as follows: Section \ref{sec:bipartite} introduces some fundamental concepts of bipartite networks and defines the properties we are interested in, such as degree distributions, clustering coefficients, and redundancy. Section \ref{sec:networks} details the features of our empirical network alongside the properties of synthetic networks and explains some challenges of modeling bipartite structures. Section \ref{sec:col_features} outlines collaborative metrics for collaboration networks, and discuss the implications of such metrics in regard to different strategies adopted by institutions when collaborating. In Section \ref{sec:conclusions}, we present our conclusion of this work.

\section{\label{sec:bipartite}Bipartite networks}
Before we look at co-patenting networks, we take the time to introduce some concepts that are relevant for the study of bipartite networks. This step is necessary in order to identify how different measures in the bipartite network are related to those in the one-mode network.

\subsection{\label{sec:degdist}Degree Distributions}

A bipartite networks is a graph $B = \{U,V,E\}$, where $U$ and $V$ are disjoint sets of $|U|$ and $|V|$ nodes, respectively, and $E = \{(u,v):u \in U, v \in V\}$ is the set of links or edges connecting nodes. We will refer to the sets $U$ and $V$ as the bottom and top partitions respectively. Nodes $u$ $\in$ $U$ can only connect to nodes $v$ $\in$ $V$ and vice-versa. No connections among nodes of the same set are allowed. Each set of nodes can have independent properties, such as the probability distribution for their node degree, or the number of nodes (system size). Hence, for the sake of notation, we have\\
\\
$k_{u}$: degree of node $u$ $\in$ $U$,\\
$d_{u}$: degree of node $v$ $\in$ $V$,\\
$P_{b}(k)$: degree distribution of bottom nodes,\\
$P_{t}(d)$: degree distribution of top nodes.\\
\\
The total number of edges $|E|$ of the bipartite graph is given by
\begin{equation}
\label{eq:totalk}	
|E| = \sum_{u\in U} k_{u} = \sum_{v\in V} d_{v},
\end{equation}
and the density of the bipartite network $B$ is
\begin{equation}
\rho_{B} = \frac{|E|}{|U| \times |V|}.
\end{equation}
Many real-world social and economic networks are naturally bipartite structured. A few examples are business networks between companies and banks \cite{souma2003complex}, scientific collaborations on journal publications \cite{newman2001scientific}, football players and clubs they have played for \cite{onody2004complex} and the famous actor-movie networks \cite{watts1998collective}.

All of the above examples make use of one of the most interesting properties of a bipartite network, its one-mode projection. 
A projection onto the nodes $U$ (a so-called bottom projection) results in a one-mode network, $G = \{U,L\}$ where node $u$ is connected to $u'$, $\{{u,u'}\}$ $\in$ $U$, only if there exists a pair of edges $(u,v)$ and $(u',v)$ $\in E$, such that $u$ and $u'$ share a common neighbor in $V$. Similarly, in a projection onto $V$ (or top projection) a node $v$ is connected to node $v'$ if they share a neighbor in $U$. 

In order to simplify, from now on we will refer to projections as bottom projections unless otherwise stated. Thus, also for the sake of notation, we have for projected networks\\
$q_{u}$: degree of node $u$ $\in$ $U$,\\ 
$P(q)$: bottom projection degree distribution.\\
The bottom nodes $u$ $\in$ $U$ have now degree $k_{u}$ in $B$ and degree $q_{u}$ in $G$. 

A projection can be built as a simple graph (only single edges are allowed between pairs of nodes in the projection), a weighted graph (only single edges, possibly weighted), or as a multigraph (multiple degenerate edges are allowed between any pair of nodes in the projection) depending on how one wishes to treat the projection of node pairs that share more than one common neighbor in the bipartite network. 
Weighted links in a projection allow one to keep track of the number of common neighbors a pair of nodes share \cite{zhou2007bipartite,fan2007effect,newman2001scientificII} in the bipartite network, while still only allowing a unique link between node pairs in the projection. In this case, the number of neighbors of $u$ in $G$ and its degree are the same. In a multigraph projection there exist parallel links between pairs of nodes call multilinks. They are another way of keeping track of common neighbors shared by pairs of nodes in the bipartite structure. However, each link in the projection contributes to the node degree, hence the degree distribution changes and degrees are no longer representative of the number of unique neighbors that nodes have in the projected network. Instead the degree of node $u$ represents, in this case, the total number of interactions such nodes have with their neighbors in the projection $G$. The degree of a node in a multigraph projection is also called the strength of the node and is given as the sum of the weights of the links connected to a node in a weighted network \cite{barrat2004architecture}. 

At this stage, an important relation can be inferred from the above definitions. Considering that node $u$ is connected to $k_{u}$ nodes $v_j\in V$ and each one of these nodes is connected to $d_{v_j}$ nodes $u'\in U$, with $j = {1,2,...,k_{u}}$, then

\begin{equation}
\label{eq:qi}
q_{u} \leq \sum_{j=1}^{k_{u}}(d_{v_j} - 1),
\end{equation}
where the expression becomes an equality if the projection is a multigraph.
The degree $q_{u}$ of node $u$ in the bottom projection is the sum of the degrees $d_{v_j}$ of all nodes it is connect to minus its own degree $k_{u}$. When the projection is a (weighted) simple graph, the expression is an equality only in the case where the bottom nodes never share more than one common neighbor from the top partition.

\subsection{\label{sec:clustering}Clustering Coefficient}

The search for methods to measure clustering in networks came with the knowledge that relations in social systems tend to form group structures, or communities. The measure which is now named as the global clustering coefficient was first coined with the idea of counting open and closed triplets using adjacency matrices of one-mode networks \cite{luce1949method}. An open triplet is a sequence of three nodes connected by two edges (a two-path) while a closed triplet is defined as three connected nodes forming a cycle, that is a triangle (three-cycle). 
The global clustering coefficient of a one-mode network is
\begin{equation}
\label{eq:global_clust}
	C = \dfrac{\text{\# closed triplets}}{\text{\# triplets}} = \frac{3\times \text{\# triangle}}{\text{\# triplets}},
\end{equation}
and represents the universal measure of clustering found in a network. The need to multiply the number of triangles by three is due to the fact that each triangle contains three closed triplets, each triplet centered in one of the nodes of triangle.

On the other hand, the local clustering coefficient is centered on a particular node $u$ and represents the level of clustering found in its immediate neighborhood \cite{watts1998collective}. The  local clustering coefficient can still be evaluated as the ratio of the number of closed triplets to the total number of triplets. In this case, however, the triplets are only centered on the node in question. An alternative way of looking at the local clustering coefficient is to find the fraction of existing links between neighbors of $u$ in the projection $G$, from the maximum possible links between them \cite{barrat2000properties,watts1998collective}. Considering $q_{u}$ as the number of first neighbors of $u$ in $G$ ($N(u) = q_u$), the total possible number of links between its neighbors is $\frac{q_{u}(q_{u}-1)}{2}$. Therefore, the local clustering coefficient is given by
\begin{equation}
	cc_{u} = \dfrac{2L_{u}}{q_{u}(q_{u}-1)},
\end{equation}
where $L_{u}=\{\{u',u''\} \subseteq N(u):~ (u',u'') \in L\}$ is the number of existing links between the $q_{u}$ neighbors of $u$. For the whole network, the average clustering coefficient is 
\begin{equation}
	\label{eq:av_cc}
	\langle c \rangle = \dfrac{1}{|U|}\sum_{u\in U}cc_{u}.
\end{equation}

The clustering metrics presented above apply to simple graphs, that is, they do not allow self-loops, directed edges, or multiple edges between pairs of nodes, and they do not take into account the weighted links in the one-mode network \cite{west2001introduction}. 

For a little over a decade, many studies have aimed to find generalizations of the clustering coefficient according to the type of network involved. Some of the new generalization propositions range from directed \cite{fagiolo2007clustering} and weighted \cite{fagiolo2007clustering,barrat2004architecture,opsahl2009clustering,ahnert2007ensemble} one-mode networks to bipartite \cite{robins2004small,zhang2008clustering,latapy2008basic,opsahl2013triadic}, multilayer and multiplex \cite{boccaletti2014structure,cozzo2015structure,battiston2014structural,baxter2012avalanche} networks. Due to the purpose of our work, we will narrow our focus to the developments related to clustering coefficients for bipartite networks.

There are no three-cycles (or triangles) in bipartite networks since no connections are allowed between nodes of the same type. The minimum possible cycle in such networks is a four-cycle, or a square. Therefore one proposed generalization for the global clustering coefficient is the ratio of the number of closed four-paths to the total number of three-paths present in the bipartite network \cite{robins2004small}, according to   
\begin{equation}
C = \frac{\text{\# 4-cycles}}{\text{\# 3-paths}} =\frac{4 \times \text{\# squares}}{\text{\# 3-paths}}\,.
\end{equation}
This is similar to the case of one-mode networks with triangles; here, we need to multiply the number of squares by a factor of four, instead of three, as there are four possible open three-paths in a square.

\begin{figure}[!ht]
\centering
\captionsetup{width=.82\linewidth}
{\includegraphics[scale=0.50]{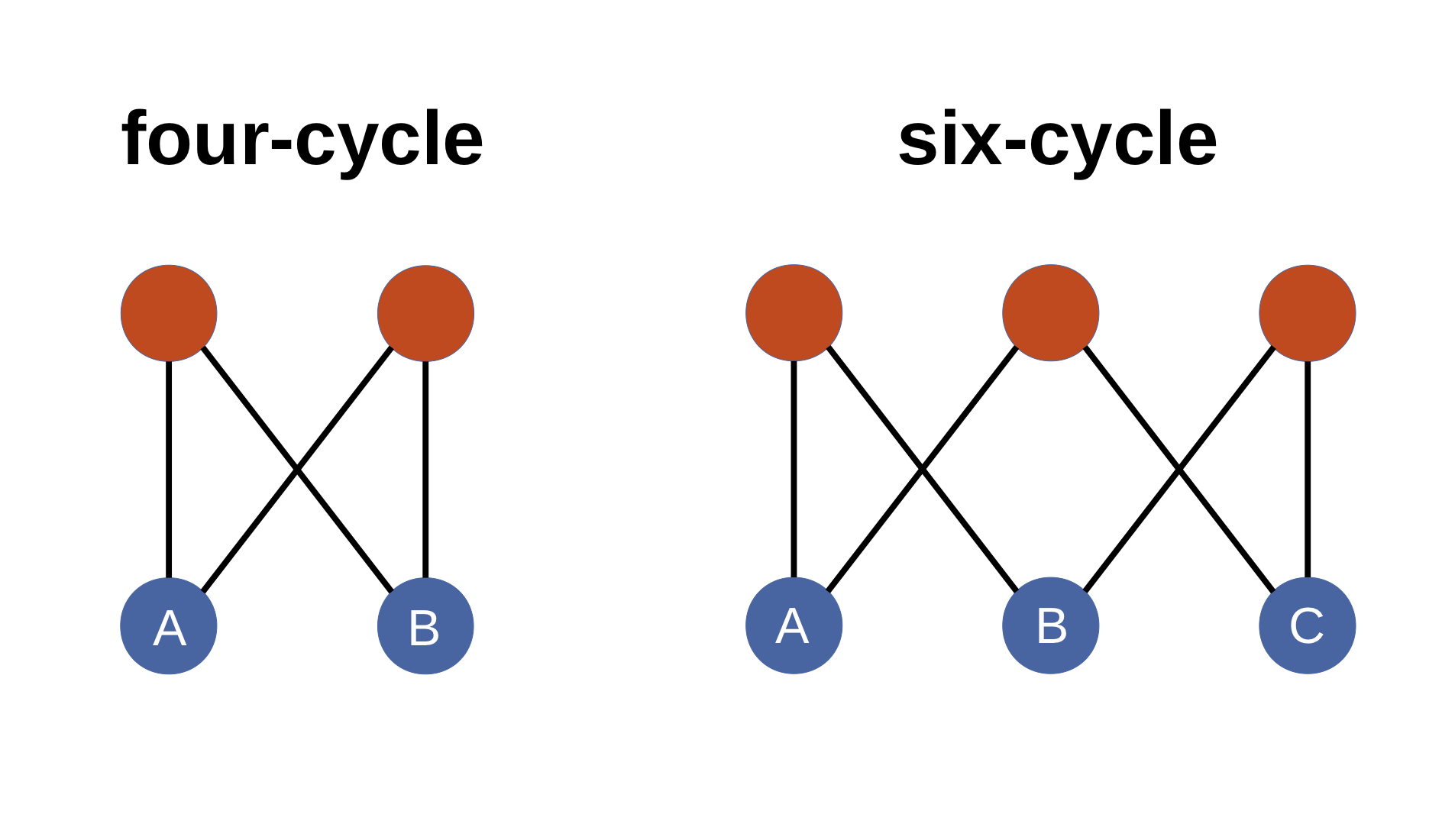}}
\caption{Schematic for four-cycle and six-cycle motifs found in bipartite networks.}
\label{fig:motifs_bip}
\end{figure}

Although this generalization is appealing, on account of being similar to that for one-mode networks, there is a recent reasoning that goes against it \cite{opsahl2013triadic}. The core of the criticism is the fact that in a square, only two agents are included, sharing two common neighbors (artifacts). Using the example of the co-patenting network, this means that two institutions are collaborating twice, filing for two different patents. Even though a square is a form of clustering in bipartite networks, it does not involve three nodes, hence there is no triadic closure among the institutions. In order to involve three agents, an open four-path is needed, connecting three institutions and two patents. To close a four-path, one more patent, with two more links, is required. This new patent connects to the first and to the last institutions of the chain, forming a six-cycle (Fig. \ref{fig:motifs_bip}). The corresponding global clustering coefficient, $C$ is therefore given by \cite{opsahl2013triadic}
\begin{equation}
\label{eq:global_bip_clust}
C = \frac{\text{\# 6-cycles}}{\text{\# 4-paths}}\,.
\end{equation}  
The local clustering coefficient of a given node $u$ is calculated using Eq. (\ref{eq:global_bip_clust}) by counting the cycles and paths centered on node $u$. The average clustering for the network  can then be obtained using Eq. (\ref{eq:av_cc}).

A different approach to quantifying clustering in bipartite networks is seen in \cite{latapy2008basic} where, instead of counting squares (or six-cycles) and three-paths (four-paths), the authors propose measuring the overlap between neighborhoods of pairs of nodes. Hence, prior to measuring the clustering coefficient of one particular node, a calculation is performed for pairs of nodes in $U$ that have at least one common neighbor in $V$. The clustering coefficient $cc_{uu'}$ for the pair of nodes $uu'$ is given by the Jaccard index
\begin{equation}
\label{eq:cc_uu}
cc_{u,u'} = \frac{|N(u) \cap N(u')|}{|N(u) \cup N(u')|}\,,
\end{equation}  
where $N(u)$ is the set of neighbors of $u$. In order to obtain the overlap clustering coefficient $cc_{u}$ for a given node $u$, we take the average of the pairwise clustering coefficients of $u$ with all its neighboring nodes, according to
\begin{equation}
cc_{u} = \frac{\sum_{u'\in N(N(u))} cc_{u,u'}}{|N(N(u))|}\,,
\end{equation}
where $N(u)$ is the set of neighbors of $u$, $N(N(u))$ is the set of next-nearest neighbors of $u$ in $U$, and $cc_{uu'}$ is the coefficient defined by Eq. (\ref{eq:cc_uu}).

In Section \ref{sec:networks}, we compare these two notions --- six-cycles and overlap --- of local clustering coefficient for bipartite networks. We look at the level of clustering in the original bipartite networks and their implications for the projections. In order to distinguish between the measures, from now on we will refer to the first as the six-cycle clustering coefficient, $cc_{u}^{6}$ and the latter as the overlap clustering coefficient, $cc_{u}^{o}$.

The notion of overlap will also be useful for explaining the redundancy coefficient (Section \ref{sec:redundancy}) --- an alternate measure, also proposed by the authors of \cite{latapy2008basic} --- that we will employ in our bipartite network analysis alongside the above measures of clustering. 

\subsection{\label{sec:redundancy}Redundancy Coefficient}

\begin{figure*}[h!]
\centering
\captionsetup{width=.82\linewidth}
{\includegraphics[scale=0.50]{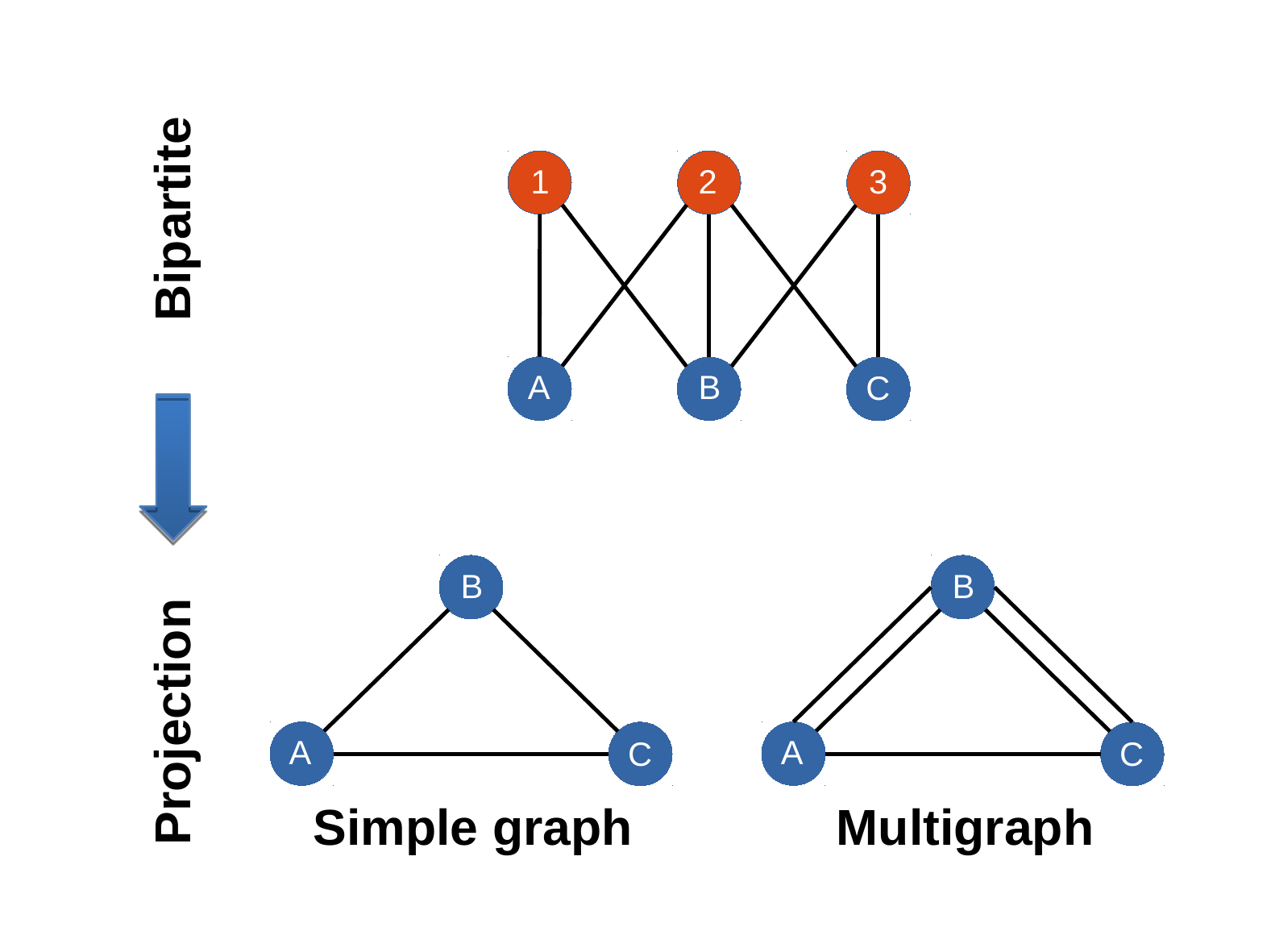}}
\caption{Schematic for the redundancy coefficient. Node 2 induces three edges in the projected network, creating the triangle $\textrm{ABC}$. Node 1 connects A and B and node 3 connects B and C. In the multigraph case, two multilinks are connecting the pair AB and the pair BC. However, in a simple graph projection, such multilinks are amalgamated into one. Then if node 1 is removed, nodes A and B would still be connected. As node 1 only induces the link $(\textrm{A},\textrm{B})$, its redundancy coefficient is $rc_{1}=1$. Following the same reasoning, $rc_{2}=\frac{2}{3}$ and $rc_{3}=1$.}
\label{fig:redundancy_drawing}
\end{figure*}

As we have seen in the previous section, the  overlap clustering coefficient of $u$ is obtained by taking the average of the pairwise clustering coefficients of $u$ and its neighbors. The notion of redundancy was introduced to create a different idea of overlapping metrics, this time with respect to a specific node \cite{latapy2008basic}. The \textit{redundancy coefficient} of a node $v$ is the fraction of links induced by it that would not disappear from a simple graph projection if $v$ was to be removed from the graph. Such links would persist in the simple graph as links induced by nodes other than $v$. In a multigraph projection, the redundancy coefficient would be the fraction of links induced by $v$ that are part of multilink connections, as shown in Fig. \ref{fig:redundancy_drawing}. The redundancy coefficient defined in \cite{latapy2008basic} is given by:

\begin{equation}
\label{eq:redundancy}
rc_{v} = \frac{|\{\{u,u'\} \subseteq N(v):~ \exists v'\neq v \text{ with } (v',u) \in E \text{ and } (v', u') \in E\}|}{d_{v}(d_{v}-1)/2}\,, 
\end{equation} 
where $N(v)$ is the set of neighbors of $v$. 

\section{\label{sec:networks}Empirical and synthetic networks}

After establishing and detailing the networks structure and metrics of our interest, we now introduce the characteristics of the empirical data, object of our study. Also, we present the methods to create the collaboration and synthetic networks.

We constructed a patent collaboration network using the OECD HAN (Harmonised Applicant Names) and RegPat databases for the European Patent Office records, edition of February 2016 \cite{patstat}. Patent applicants are the agents (or bottom nodes) and patents numbers are the artifacts (or top nodes). The bipartite network comprises a total of 3,098,113 nodes, with 313,769 applicants (agents) or bottom nodes and 2,784,344 patents (artifacts) or top nodes. A summary of the network properties are detailed in Table \ref{table:net_summary}. We compared the properties of the empirical collaboration network with those from synthetic analogues that we created using two models. 

\begin{table}[!ht]
\captionsetup{width=.82\linewidth}
\caption{Summary of the basic properties of the empirical bipartite network created from the OECD HAN and RegPat datasets \cite{patstat}, comprehending 40 countries and dating back to 1978.}
\label{table:net_summary}
\centering
\begin{tabular}{lr}
\hline
Number of patents $|V|$                   & 2,784,344 \\
Number of institutions $|U|$              & 313,769   \\
Number of links $|L|$                     & 2,970,438 \\
Density of the network $\rho_{\textrm{B}}$ & $\num{3.4e-6}$ \\
Highest top degree $d_{\textrm{max}}$     & 61 \\
Mean top degree $\langle d \rangle$    & 1.07 \\
Highest bottom degree $k_{\textrm{max}}$  & 37,992 \\
Mean bottom degree $\langle k \rangle$ & 9.47 \\
\hline
\end{tabular}
\end{table}

First, we used a random model in which we preserved the number of nodes in each node set and the total number of links in the network and randomly rewired them. This is a bipartite variant of the Erd\H{o}s and R\'{e}nyi $G(|U|,|L|)$ model \cite{erdds1959random}, where $G$ is a one-mode network, with $|U|$ nodes and $|L|$ links. This model creates random bipartite networks, $B_{\textrm{ER}}(|U|,|V|,|E|)$, with $|U|$ bottom nodes, $|V|$ top nodes and $|E|$ links.

Secondly, we used a configuration model \cite{bender1978asymptotic,bollobas1980probabilistic,filho2018degree}, which is also a random model but with a specified degree sequence for each node set. Here, we used the degree sequence found in the empirical network, performing a random rewiring process to obtain our synthetic networks, $B_{\textrm{CM}}$. From now on we will refer to the latter as the configuration model and to the first as the ER model.

\begin{figure*}[!ht]
\centering
\captionsetup{width=.82\linewidth}
\subfloat{\label{fig:emp_dd} \includegraphics[scale=0.33]{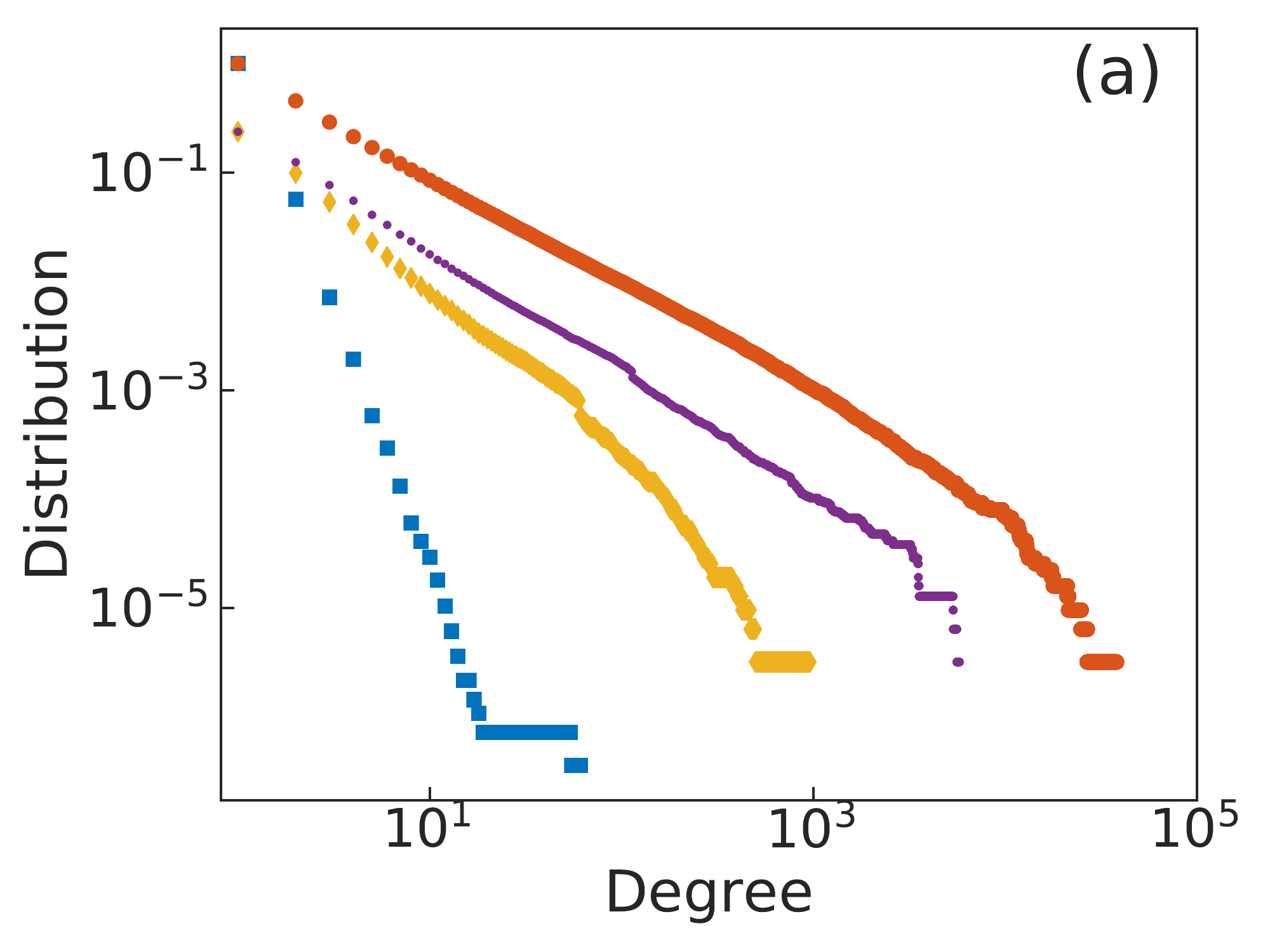}}
\subfloat{\label{fig:cm_dd} \includegraphics[scale=0.33]{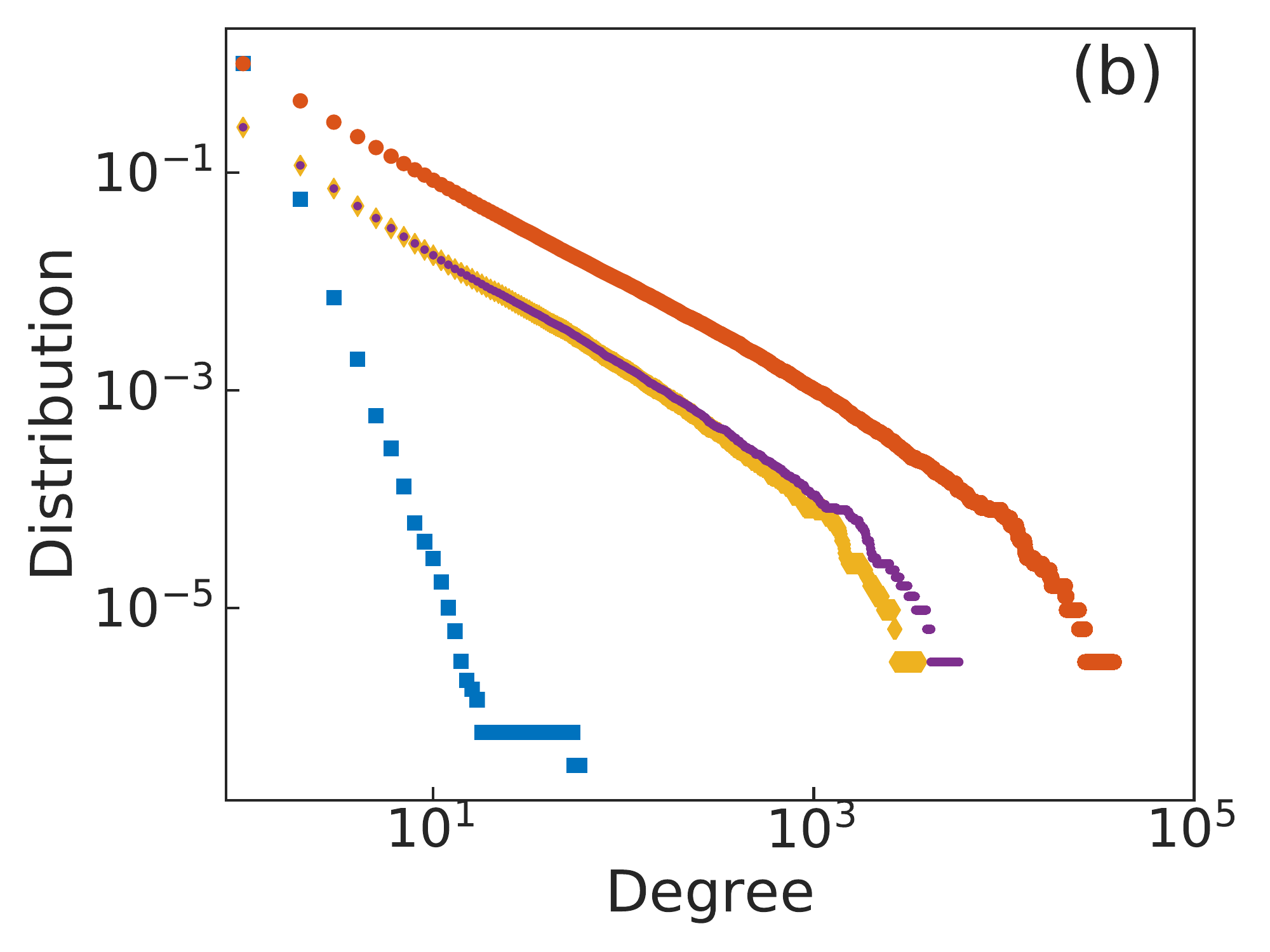}} \hfill
\subfloat{\label{fig:r_dd} \includegraphics[scale=0.33]{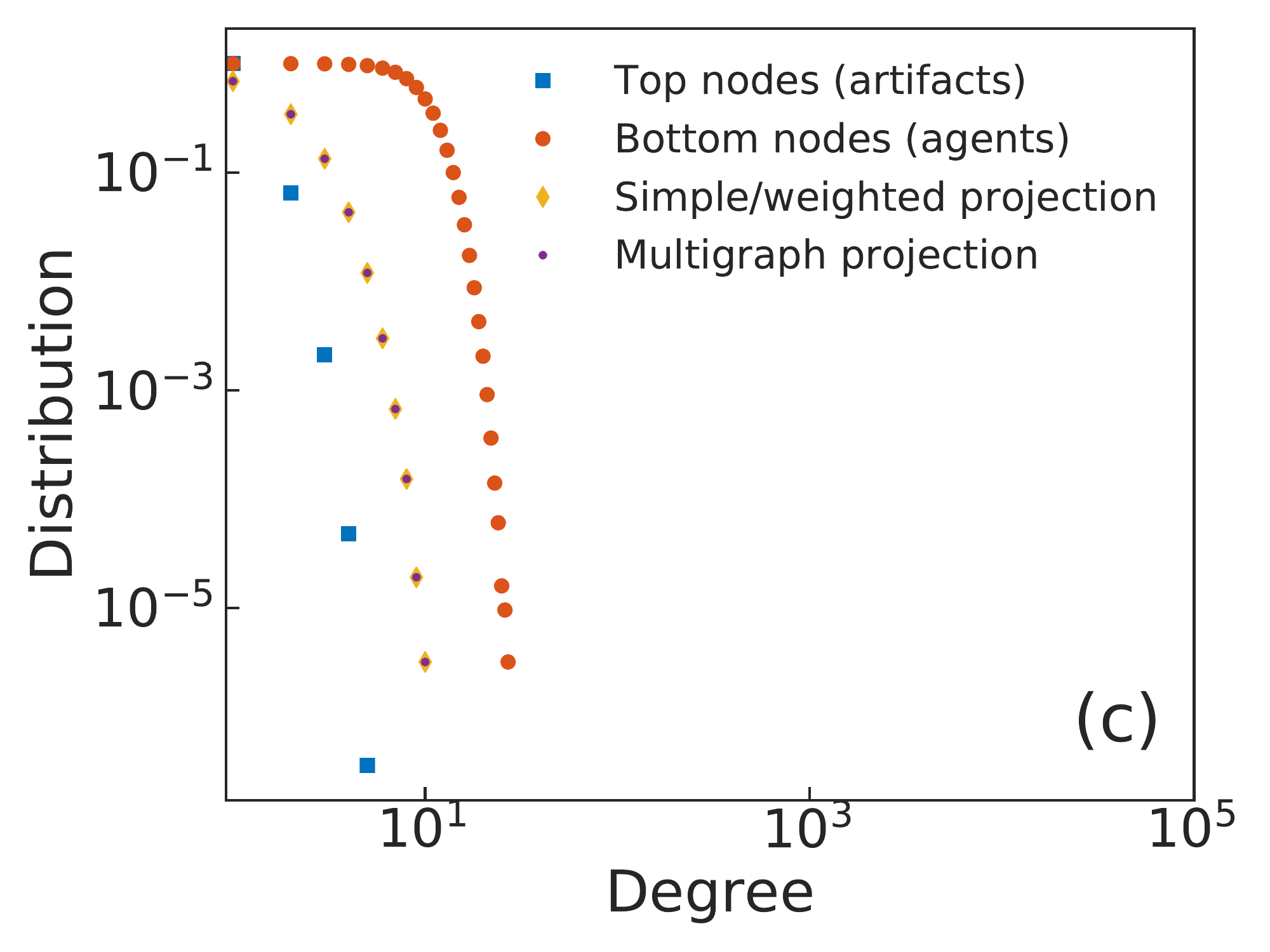}}
\caption{Degree distributions (CCDF --- complementary cumulative distribution function) of bipartite and projected networks. (a) Empirical co-patenting network: the difference between the simple/weighted graph projection and the multigraph projection shows that there are a large number of multilinks in the multigraph due to pairs of applicants co-filing for multiple patents. (b) Configuration model: distributions are similar to those of the empirical network, except the simple/weighted graph projection. Although the degree sequence of the bipartite network is kept, the link rewiring decreases the number of top nodes (patent applications) that are shared by pairs of bottom nodes (applicants). (c) ER network: Distributions are peaked and there is no significant difference between the projections. Degree distributions differ markedly from those of the empirical and configuration networks.}
\label{fig:degree_dist}
\end{figure*}

The first network property that we look at is the degree distribution. For the projection of the empirical network, the degree distribution of the simple/weighted graph is significantly different from the node strength distribution (i.e. the degree distribution of the multigraph projection). This is easily explained by the fact that the empirical bipartite network contains many agents that share more than one common neighbor from the opposite set of nodes (i.e. institutions that apply for patents together more than just once). This is the first of several pieces of evidence that we will see shortly, showing the tendency of some institutions to collaborate more with their usual partners, leading to a low diversity of collaborators. On the other hand, for the configuration model, degree and node strength distributions are fairly similar to each other, as we can see in Fig. \ref{fig:degree_dist}. This is due to the fact that links in the model are randomly rewired, decreasing the number of common neighbors for an arbitrary pair of bottom nodes.

The magnitude of the difference between degree and node strength distribution can be explained in terms of the density of the network. The maximum possible number of links  a projection can have is given by	
\begin{equation}
\label{eq:lp_max}
|L|_{\textrm{max}} = {\sum_{v=1}^{|V|} \frac{d_{v}(d_{v}-1)}{2}}\,.
\end{equation}	

In contrast, a simple one-mode network has at most $\frac{|U|(|U|-1)}{2}$ links. 
Then, the density of a multigraph projection, $\rho_\textrm{P}$, due to its original bipartite structure is given by	
\begin{equation}
\label{eq:rho_p}
\rho_\textrm{P} = \frac{\sum_{v=1}^{|V|} d_{v}(d_{v}-1)}{|U|(|U|-1)}\,,
\end{equation}
where $|U|$ is the number of agents or bottom nodes and $|V|$ is the number of artifacts or top nodes.
This is also the average probability of two institutions being connected (i.e. the average probability of one link connecting two institutions in a multigraph projection). The actual probability depends also on the degree of these institutions in the bipartite network. Hence, the average probability of finding two links connecting a single pair of nodes in the projected network of the configuration model is the square of Eq. (\ref{eq:rho_p}). Due to the very sparse nature of our empirical bipartite network ($\rho_{\textrm{B}}=\num{3.4e-6}$), the probability of a rewired link connecting a pair of nodes that are not already connected in the projected network of the configuration model is much bigger than the probability of creating an additional connection for two already connected nodes. As mentioned, this is the average probability, as the pairwise connection probability depends on the degrees of both bottom nodes in the bipartite network, according to
\begin{equation}
P[(u,v)] = \frac{\binom{|E|}{k_{u}}-\binom{|E|-d{v}}{k_{u}}}{\binom{|E|}{k_{u}}}\,,
\end{equation}
and
\begin{equation}
P[(u',v)|(u,v)] = \frac{\binom{|E|-1}{k_{u'}}-\binom{|E|-d{v}-1}{k_{u'}}}{\binom{|E|-1}{k_{u'}}}\,.
\end{equation}

One can think of the average probability --- given by Eq. (\ref{eq:rho_p}) --- as the probability of a pair of nodes having more than one common neighbor in the bipartite network during the rewiring process. Therefore, the probability of the rewiring process linking together nodes that were previously disconnected in the empirical case is much larger than the probability of preserving the total number of common neighbors of any pair of nodes in the network.

When we look at the degree distributions of the bipartite structure, using the ER model, we see they are peaked, following a Poisson distribution \cite{newman2001random, filho2018degree}. As expected, the same distribution shape is observed for both the simple/weighted graph and multigraph projections (Fig. \ref{fig:r_dd}) \cite{filho2018degree}. 

We also calculated the degree assortativity $r$ (the degree-degree correlation, or the tendency of nodes being connected to others with a similar degree) of the co-patenting network, according to  Eq. (2) of \cite{newman2003mixing}. Positive values of $r$ indicate networks with degree assortativity (degree-degree correlation), while negative values indicate degree dissortativity (degree-degree anti-correlation), and $r=0$ indicates no degree-degree correlation. The structure of the empirical co-patenting network, unlike most social networks, does not exhibit degree assortativity, with coefficient $r=0.0046$. One reason for this is the fact that the bottom degree distribution (the number of patents per applicant) is much more right-skewed than the top degree distribution (the number of applicants per patent). Degree assortativity is more common in projected networks with a broader degree distribution of the top nodes \cite{filho2018degree,vasques2019transitivity}. That is, the lack of large collaborations for patent applications contributes to giving a neutral degree-assortative network.

\begin{figure*}[h!]
\centering
\captionsetup{width=.82\linewidth}
{\includegraphics[scale=0.5]{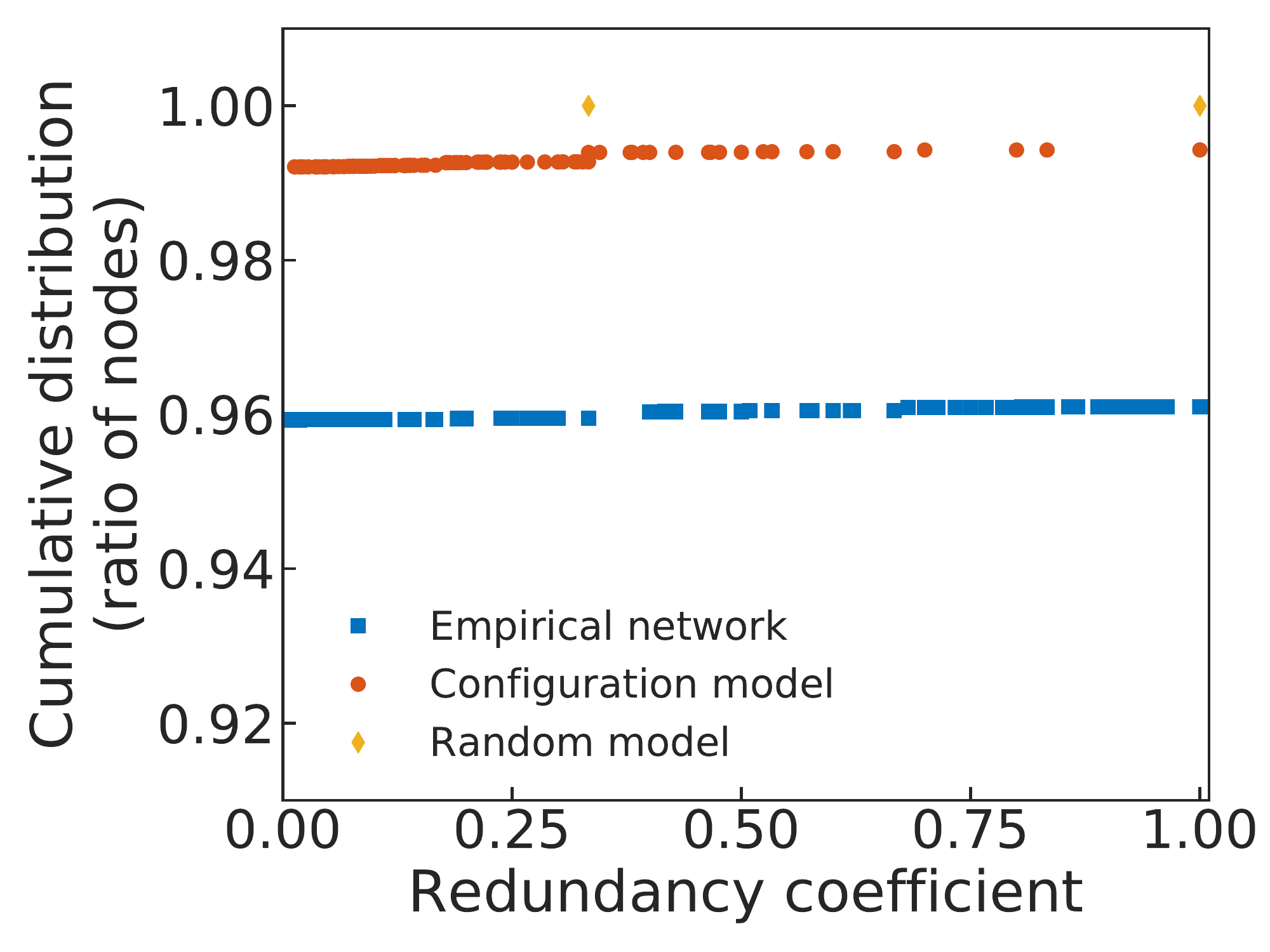}}
\caption{Redundancy coefficient distributions for empirical and synthetic networks using random and configuration models. The $y$-axis indicates the fraction of nodes that have up to the corresponding amount of redundancy. That is, for the empirical network around $96\%$ of the nodes have $rc_{v}=0$, a very small fraction of nodes has $0 \leq rc_{v} \leq 1$, and around $4\%$ are fully redundant. The rewiring process reduces redundancy. The configuration model still presents some redundancy due to fact that the degree sequence is preserved. Redundancy is practically absent with the random model (note the presence of only two markers close to the $y=1.00$ line, with corresponding $rc\approx0.35$ and $rc\approx1.00$).}
\label{fig:redundancy}
\end{figure*}

Moving forward, when we look at the bipartite network, both ER and configuration models break the redundancy structure found in the original empirical network. We notice that the empirical network does not have many top nodes with  non-zero redundancy. However, most of top nodes that do have non-zero redundancy are completely redundant ($rc_{v}=1$). This corresponds to about $4\%$ ($110,000$) of all top nodes, as shown in Fig. \ref{fig:redundancy}. This is another indication that many institutions collaborate repeatedly on patents applications with former collaborators. For a top node (patent) to have redundancy $rc_{v}=1$, there exists another top node (or top nodes) connecting this same set of collaborating institutions. 

Such highly redundant top nodes have degree $d_{v}\geq2$, so the rewiring process has the potential of breaking over $110,000$ redundant links. 
It is worth noticing, as stated before, that the proportion of those broken links due to the rewiring process depends on the degree distribution. That is why we can see in Fig. \ref{fig:redundancy} that the configuration model preserves some redundancy while, in the ER model, redundancy is practically absent.

\begin{figure*}[h!]
	\centering
    \captionsetup{width=.82\linewidth}
	{\includegraphics[scale=0.50]{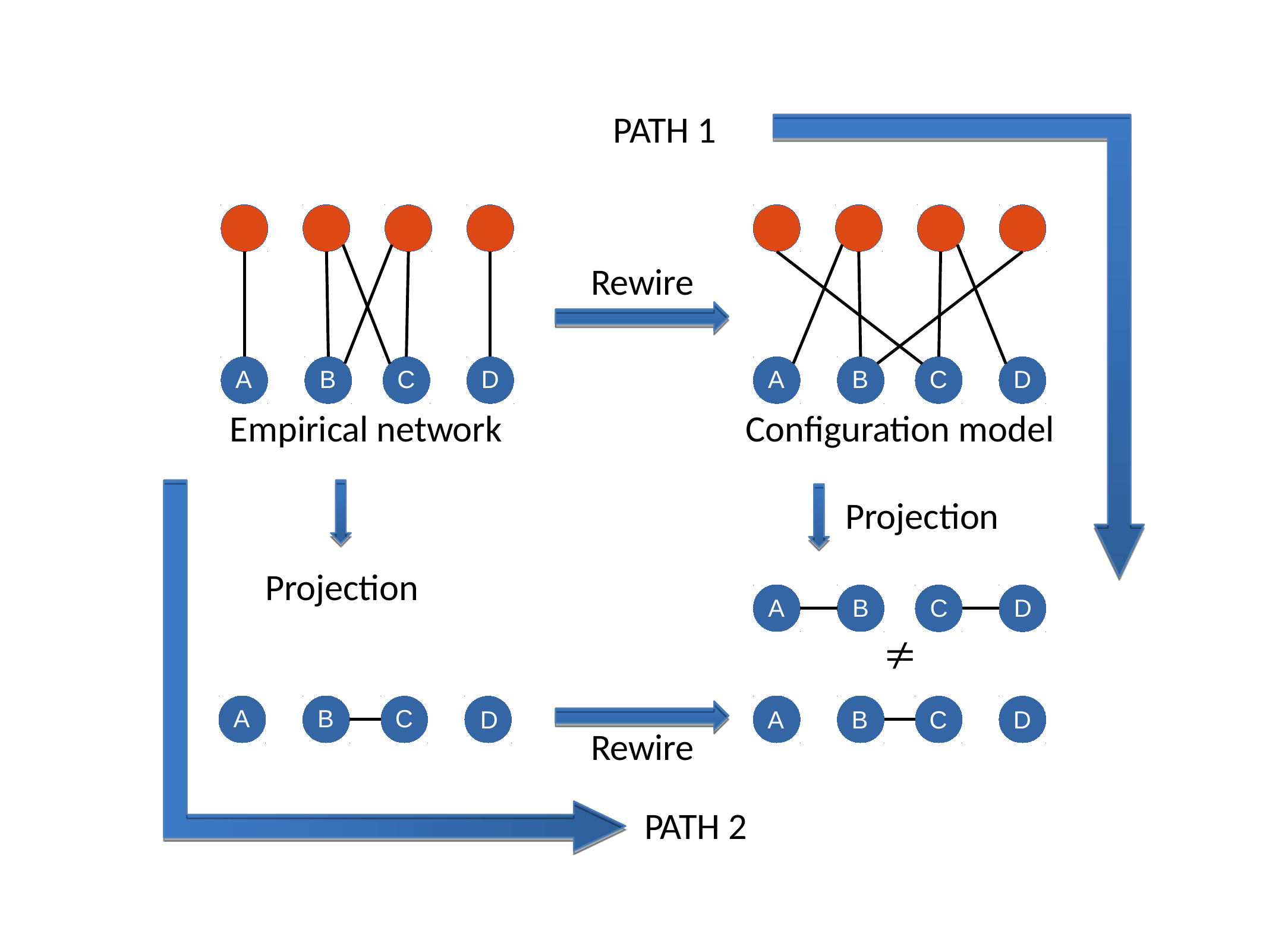}}
	\caption{Schematic showing that modeling one-mode projected networks is a non-commutative process. In fact, path 2 is a misleading way most often found in studies of models for one-mode networks from original bipartite structures. Note that as the configuration model preserves the degree sequence, in this toy network the rewiring process results in the same network configuration of the projected network, following path 2.}
	\label{fig:path1_path2}
\end{figure*}

Differences in the degree distributions between one-mode projected networks of empirical and synthetic bipartite networks indicate a non-trivial structure --- not driven only by chance --- found in the empirical bipartite network linking institutions to patents. Since the configuration model of the bipartite structure fails to reproduce the degree distribution of the simple graph projection, we turn our attention to a different approach when modeling a projected network. Instead of simulating first, keeping the degree sequence of the empirical bipartite network and then building the projection, we take another path. We first create a projected network and secondly apply the configuration model (as well as the ER model) using the empirical projected network features in order to rewire and create synthetic networks. Both processes, here called path 1 and path 2, are shown in Fig. \ref{fig:path1_path2}.  

\begin{figure*}[!ht]
\centering
\captionsetup{width=.82\linewidth}
\subfloat{\label{fig:diff_dd} \includegraphics[scale=0.33]{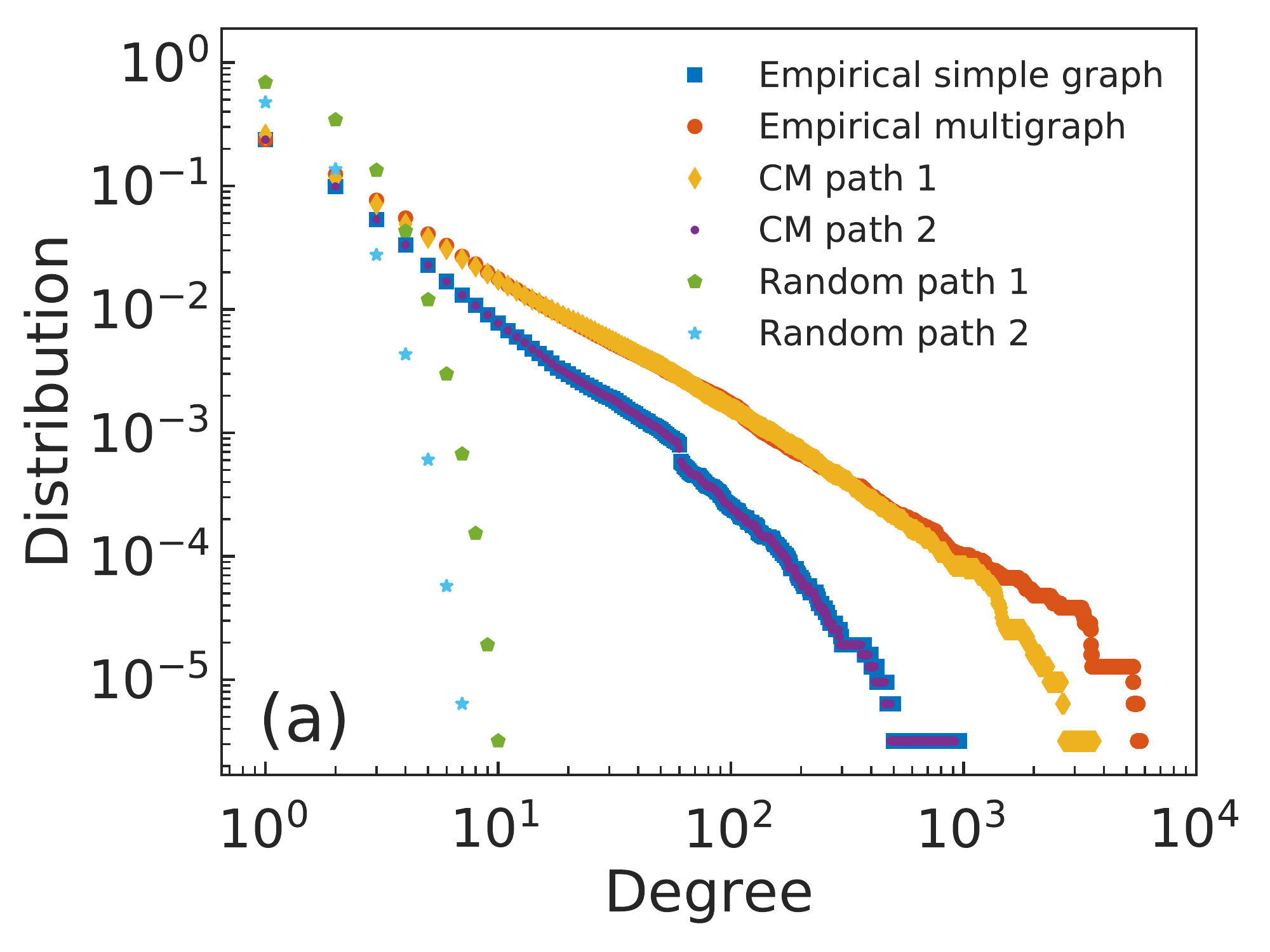}}
\subfloat{\label{fig:diff_cc} \includegraphics[scale=0.33]{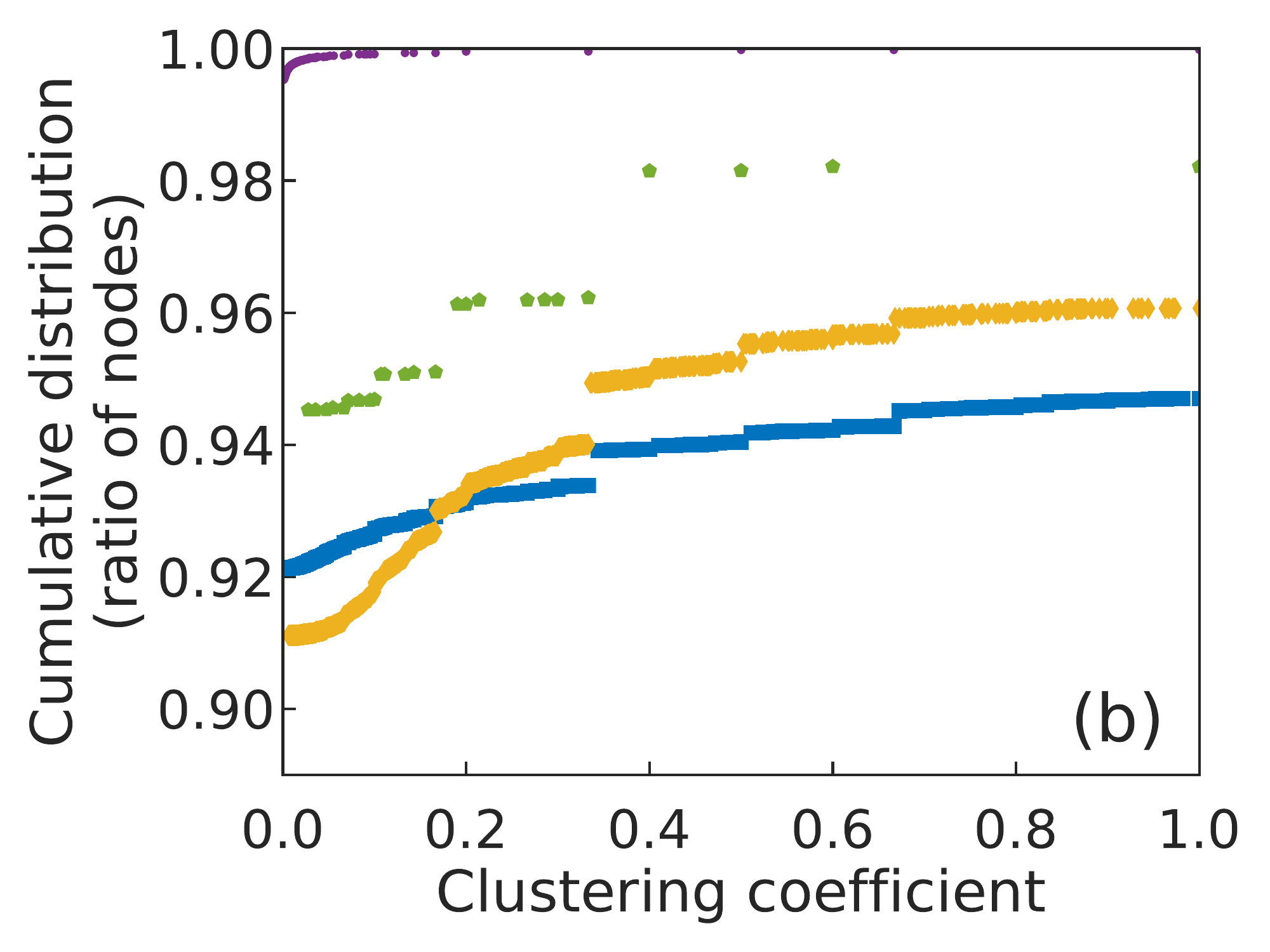}} \hfill
\caption{Statistical properties for the empirical projected network compared to synthetic projected networks through paths 1 and 2 --- see Fig. \ref{fig:path1_path2}. (a) Degree and (b) projected clustering coefficient. Path 2 is a misleading process for modeling one-mode projections of original bipartite networks. It is worth noticing that some of the clustering structure is kept even with the ER model through path 1. This is due the particular characteristic of bipartite networks in which top nodes induce complete subgraphs in the projection.}
\label{fig:diff_paths}
\end{figure*}

By following path 2 we guarantee degree distributions are the same for the empirical and the synthetic projected networks, however, this does not hold true for other properties. In fact, although this path is commonly used for modeling projected networks, it is a misleading process, as it has been noticed in \cite{tarissan2013towards}. 
Figure \ref{fig:diff_paths} compares the outcomes for the distribution of degrees and clustering coefficients of the network when modeling using both paths.

As one can see in Fig. \ref{fig:diff_cc}, the clustering structure in the projected network is completely lost when we simulate using path 2. On the other hand, some of it is kept even when path 1 is followed using the ER model. Simulating a bipartite network using the random model and then building a projection creates a one-mode network with some level of clustering due to a very particular characteristic of bipartite networks. Every top node induces a clique --- a complete subgraph --- in the projection. For instance, if a node $u$ with degree $k_{u}=1$ is connected to a node $v$ with degree $d_{v}=4$, node $v$ will create a complete subgraph with four nodes and six links, containing $u$. It will have degree $q_{u}=3$ and clustering $cc_{u}=1$. In general, even a poorly connected bottom node $u$, connected to a high degree top node $v$, will end up highly connected in the projected network. A highly connected top node, let us say with degree $d_{\textrm{max}}$, will create a fully connected subgraph with $d_{\textrm{max}}$ nodes, each with degree $q = d_{\textrm{max}}-1$, and clustering $cc=1$.

For this reason the configuration model, through path 1, also produces a more highly clustered projection than the ER model. As we preserve the degree sequence, we also preserve top nodes with higher degrees as we saw in Fig. \ref{fig:degree_dist}. Such nodes will create larger fully connected subgraphs with high clustering when the projection is built. Since path 2 destroys the clustering structure observed in the projection of the bipartite network it is a poor choice for generating synthetic networks and hence we discard it from further analysis.

Even with path 1, the level of clustering produced still does not perfectly mimic the original clustering. Again, this is an example of the non-trivial structure of the bipartite empirical network. Since redundancy is related to four-cycle motifs in the bipartite network $B$, it represents interactions between a pair of nodes only, which will induce dyads in the projection $G$. To create new triangles, other than the ones induced by top nodes with degree $d_{v}\geq3$, redundancy (four-cycle motifs showing recurrent collaboration between two institutions) is not sufficient.

\begin{figure*}[h!]
\centering
\captionsetup{width=.82\linewidth}
{\includegraphics[scale=0.50]{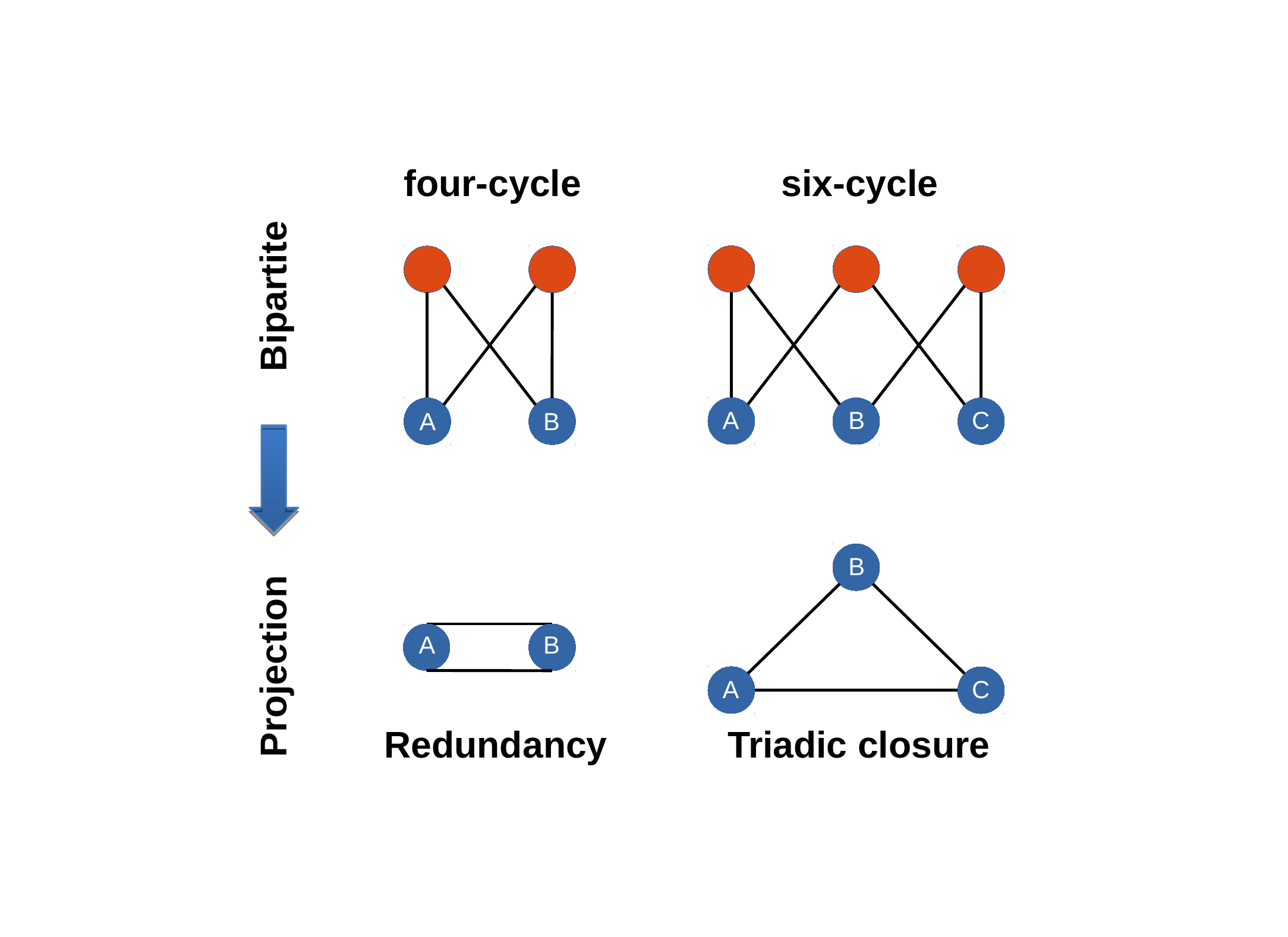}}
\caption{Schematic for four-cycle and six-cycle motifs found in bipartite networks. Other motifs and bigger cycles can also be found, but we draw our attention to these two specifically, due to their effect on one-mode projections. They are special because they create redundancy (four-cycle) and transitivity (six-cycle) in the bipartite structure, which are related to the degree distribution and clustering, in that order, of projected networks.}
\label{fig:motifs}
\end{figure*}

In order to address the formation of such triangles, we use the six-cycle clustering, in the bipartite network, which results in triadic closure in the projected network \cite{opsahl2013triadic}. This is reminiscent of the formal definition of transitivity involving three nodes,  that is if nodes $u$ and $u'$ share an artifact and nodes $u'$ and $u''$ share another artifact, then it is expected that nodes $u$ and $u''$ will share a third artifact (Fig. \ref{fig:motifs}). We can think of the clustering created due to the top nodes degree as a base level of clustering, as analytically showed in \cite{guillaume2006bipartite}, to which the transitivity due to six-cycles in the bipartite network is added. For the co-patenting network, this means two different ways of building up communities of knowledge transfer. High levels of clustering in combination with redundancy can create dense local clusters that help to increase the capacity of information transmission between the institutions. On the other hand, high levels of redundancy between the same institutions can be a limiting factor for innovation, if such institutions have restricted access to other information sources \cite{schilling2007interfirm}.  

\begin{figure*}[!ht]
\centering
\captionsetup{width=.82\linewidth}
{\includegraphics[scale=0.45]{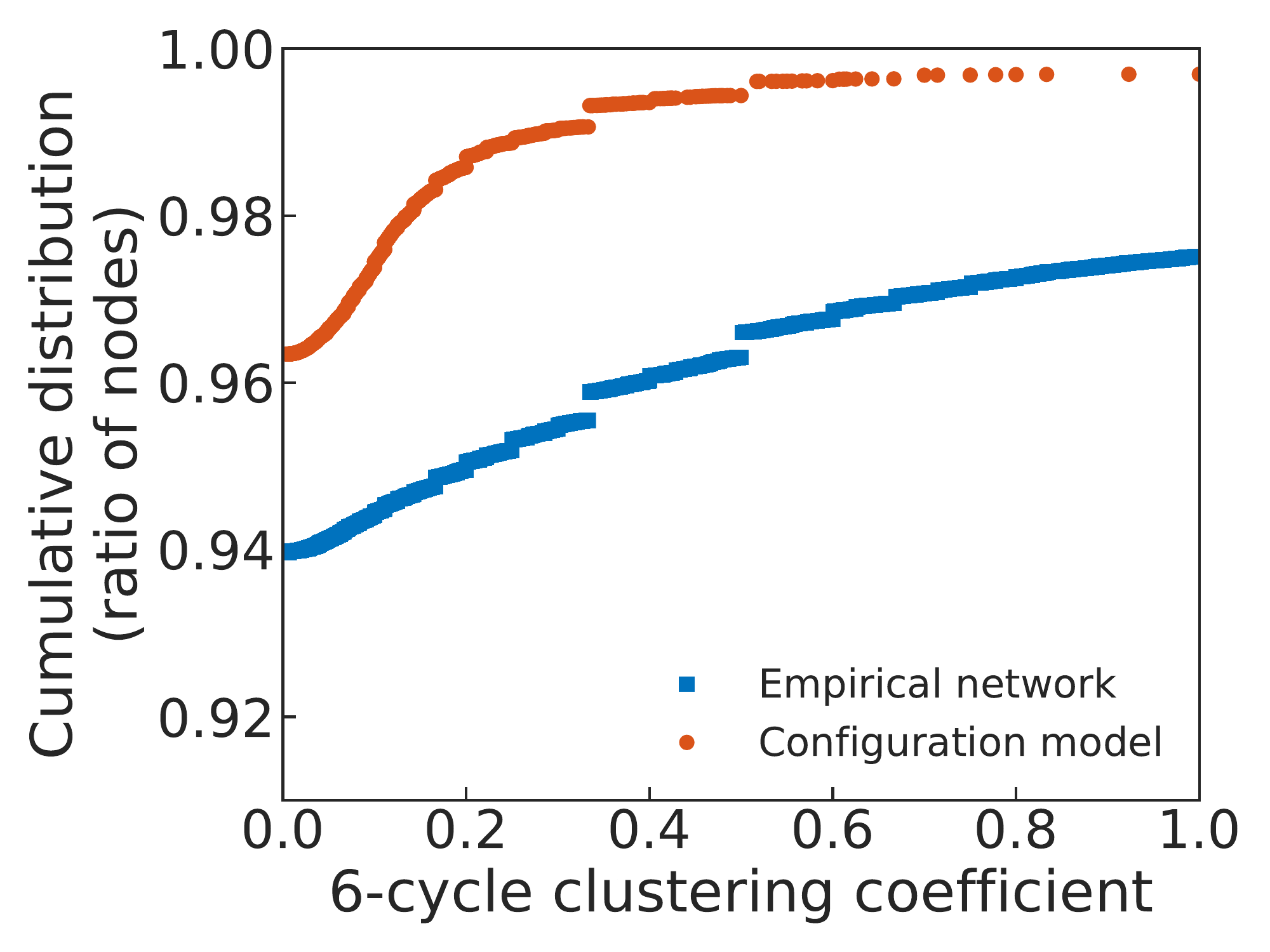}}
\caption{Six-cycle clustering coefficient distribution for empirical and synthetic networks. Similar to the case for redundancy, six-cycles are broken with the rewiring process. Some are kept in the configuration model but the six-cycle clustering is nearly zero in the random model, not even appearing in the picture.}
\label{fig:clustering}
\end{figure*}

The distributions for the six-cycle clustering for the empirical and the synthetic bipartite networks show a pattern similar to the redundancy coefficient as we can see in Fig. \ref{fig:clustering}. First, the empirical network has a small percentage, of around $6\%$ of bottom nodes presenting a non-zero six-cycle clustering coefficient. However, nearly $2.5\%$ of all agents have the six-cycle clustering coefficient $cc^{6}_{u}=1$. Second, the configuration model retains some level of clustering but much less than the empirical network. This is the rewiring process again breaking the structures found in the original empirical network, which leads to six-cycle clustering nearly being absent for the random model.

\begin{figure*}[h!]
\centering
\captionsetup{width=.82\linewidth}
{\includegraphics[scale=0.45]{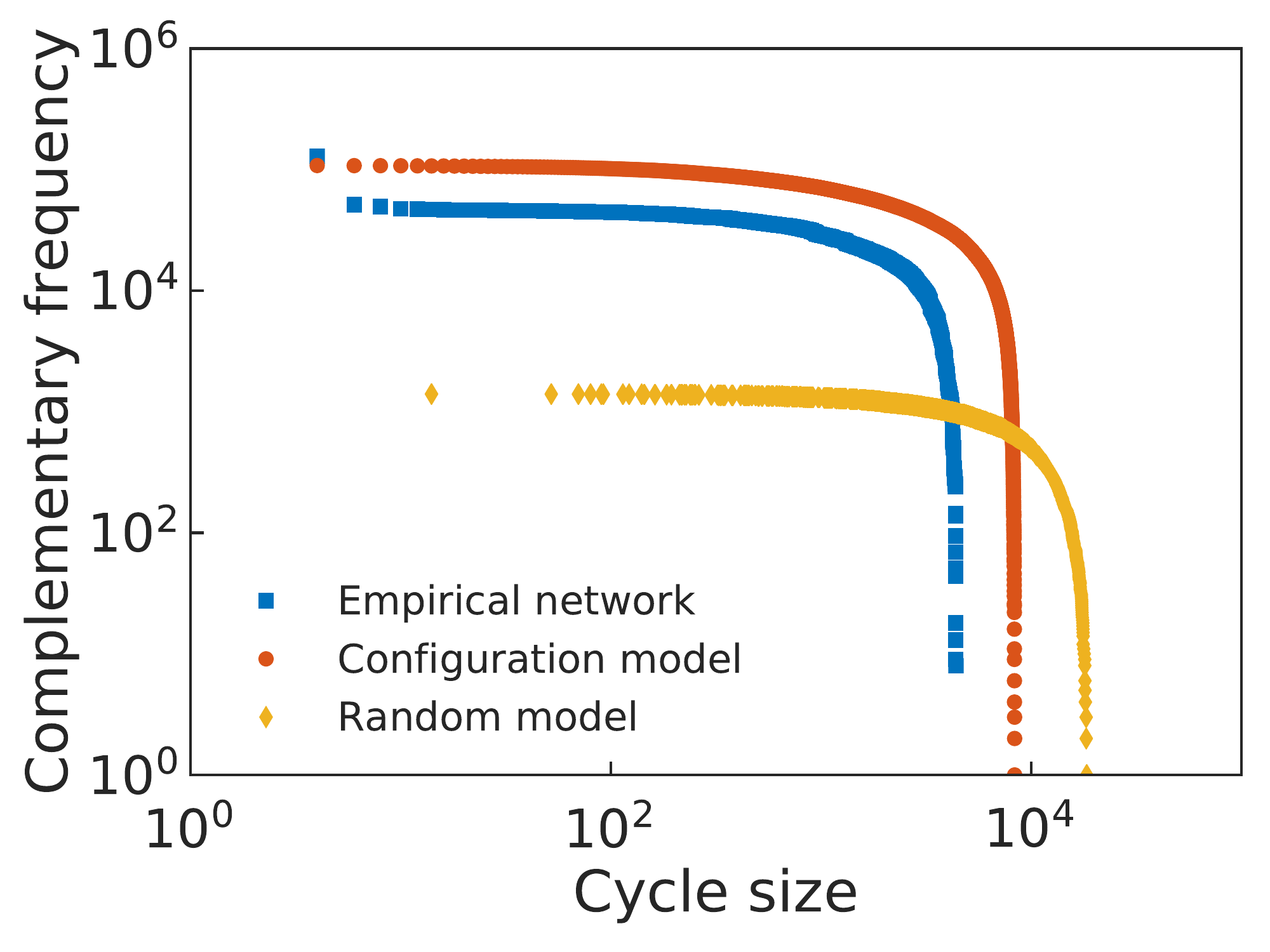}}
\caption{Frequency of sizes of cycles in the cycle basis of the empirical network and of both configuration and random models. The total number of such cycles is similar for the empirical network and configuration model due to both having the same degree sequence. However, the rewiring process breaks most of the small structures (four- and six-cycles motifs). For the random model case, such structures are almost completely broken.}
\label{fig:cycles_dist}
\end{figure*}

\begin{figure*}[h!]
\centering
\captionsetup{width=.82\linewidth}
{\includegraphics[scale=0.45]{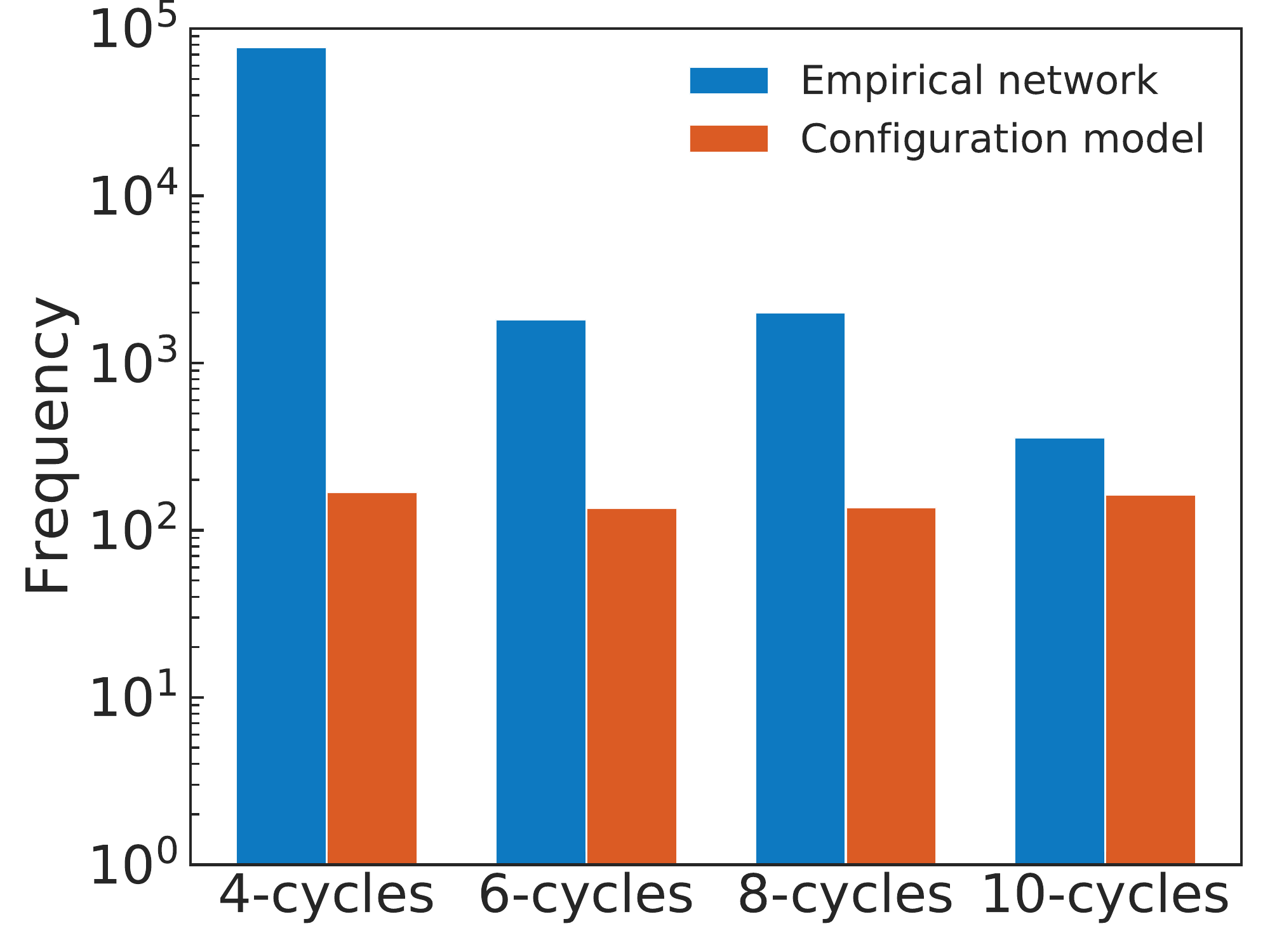}}
\caption{Frequency of small cycles going from size four to 10, in the cycle basis of the empirical network and of the configuration model. These structures are nearly absent in the random model. The non-trivial structure of real-world bipartite networks is related to small structures like the four- and six-cycle motifs}
\label{fig:cycles}
\end{figure*}

It becomes clear that small cycle motifs are of significance for the structure of real-world bipartite networks. Even though the total number of cycles  in the empirical bipartite network and the synthetic analogues using the configuration model are quite similar, the number of small cycles are very different. This difference gets smaller as the size of the cycles increases (Figs. \ref{fig:cycles_dist} and \ref{fig:cycles}). The rewiring process breaks most of these small structures and creates new larger cycles. Synthetic networks from the ER model have a much smaller number of cycles and larger cycles than observed in the empirical network and the configuration model. This behavior indicates that small structures are indeed non-trivial features of real-world bipartite collaboration networks. This reveals that information can be concentrated in specific local structures in the network, in accordance to what is discussed in \cite{schilling2007interfirm}. 

It is also worth noting that Figs. \ref{fig:cycles_dist} and \ref{fig:cycles} present the number of cycles in the \textit{cycle basis} of the graphs. The cycle basis is the set of simple cycles --- cycles where neither nodes nor edges are repeated --- from which it is possible, through combinations, to create every other cycle in the network. 

\section{\label{sec:col_features}Discussion: Collaboration properties}

Having a collaboration network as an object of study, one may ask questions like how collaborative are the agents of this network? Do they have collaborators for most of the artifacts they are connected to? Are they collaborating with many or just a few of the other agents in the network? What is the level of collaboration found in networks? We use this section to discuss collaboration properties of the network and agents using network metrics.\\

\noindent
\textbf{\textit{The collaboration density:}} is a global network property, in which we can use links of the bipartite network as a measure of collaboration intensity. Collaboration density is the density of the network subtracting links that do not represent a collaboration (i.e. those links that connect to top nodes with degree $d=1$) from the total number of links, $|E|$. Considering a network with $|V|$ artifacts and $|U|$ agents, the collaboration density is
\begin{equation}
\rho_\textrm{c} = \frac{\sum_{v}^{|V|}d_{v} - \delta_{d_{v},1}}{|V| \times |U|} = \frac{|E|-\sum_{v}^{|V|}\delta_{d_{v},1}}{|V| \times |U|}\,,
\end{equation}
where $\delta$ is the Kronecker delta. We subtract links to artifacts with $d_{v}=1$, because there is no collaboration in those cases. 
The advantage of using $\rho_\textrm{c}$ instead of $\rho_\textrm{p}$ (Eq. (\ref{eq:rho_p})), is the fact that the latter may be biased by the complete subgraphs induced in the projected networks by the high degree artifacts (patents). Our empirical bipartite network is quite sparse with density $\rho_{\textrm{B}}=\num{3.4e-6}$. As the top degree distribution is peaked, the multigraph projection is also sparse, yet denser than the bipartite network, with $\rho_{\textrm{P}}=\num{4.7e-6}$. In turn, the vast majority of the 2,784,344 patents in our data (2,636,834 patents) have $d_{v}=1$. Such patents are excluded from the calculation of the collaboration density that, for our case, has value $\rho_{\textrm{c}}=\num{3.9e-7}$. That is, the co-patenting network is not only very sparse but also presents low overall levels of collaboration.     \\

\textbf{\textit{The collaboration ratio:}} takes into account the numbers of collaborations relative to the total number of links in the network, according to

\begin{equation}
\label{eq:cratio}
R = \frac{\sum_{v}^{|V|}d_{v} - \delta_{d_{v},1}}{\sum_{v}^{|V|}d_{v}} = \frac{|E|-\sum_{v}^{|V|}\delta_{d_{v},1}}{|E|}\,.
\end{equation}
It gives the fraction of artifacts (patents) that represent a collaboration, in this case $R=0.12$. The ratio can be either a global property, as in the above equation, or a local property for a specific bottom node $u$. In the latter case, the ratio is just the proportion of artifacts associated with node $u$ that are part of a collaboration. Equation (\ref{eq:cratio}) then becomes
\begin{equation}
R_{u} = \frac{k_{u}-\sum_{v_{u}}^{k_{u}}\delta_{d_{v_{u}},1}}{k_{u}}\,.
\end{equation}
For our empirical network, $k_{u}$ is the number of patents an institution holds and $R_{u}$ is the fraction of patents such institutions share with one or more other institutions. The local collaboration ratio is the complementary of the monopoly coefficient \cite{tackx2015revealing}, which measures the proportion of top nodes with degree $d=1$ among the $k_u$ neighbors of $u$. 

\begin{figure*}[h!]
\centering
\captionsetup{width=.82\linewidth}
\subfloat{\label{fig:avshared_bipdegree} \includegraphics[scale=0.33]{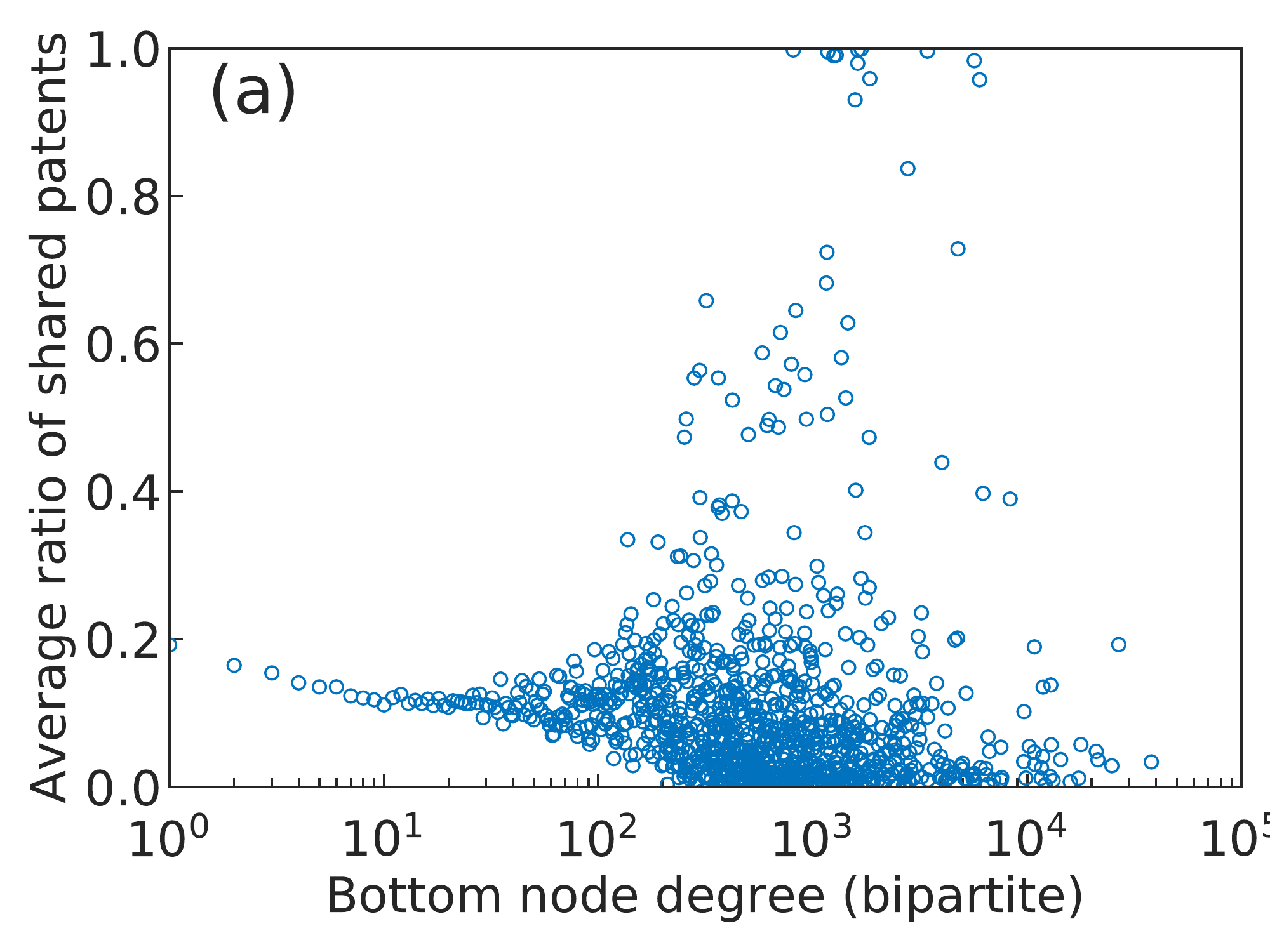}}
\subfloat{\label{fig:avshared_bipdegree_cm} \includegraphics[scale=0.33]{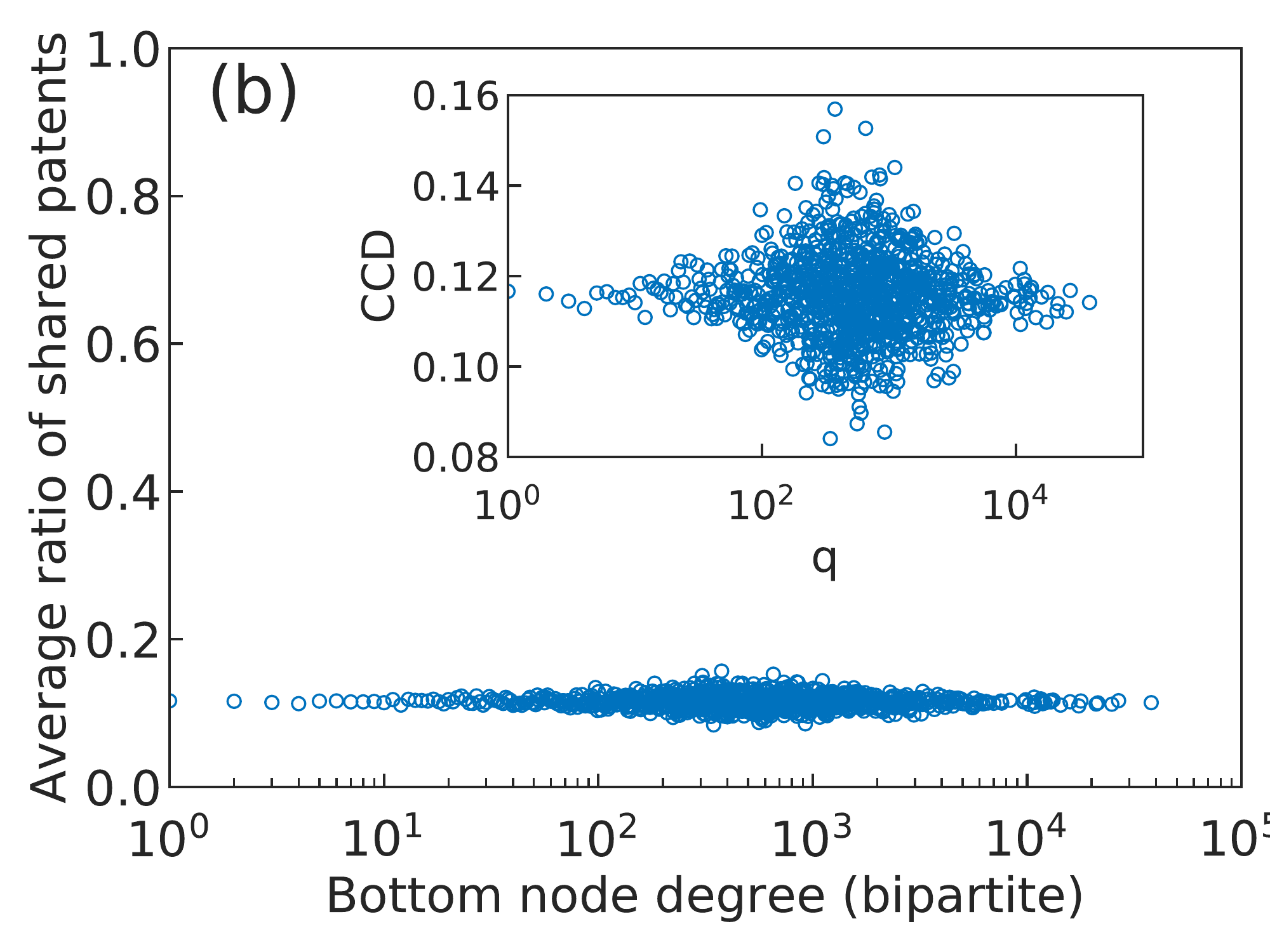}}
\caption{Correlations between average collaboration ratio (average ratio of shared patents) and bipartite degree (number of patents) for (a) empirical network and (b) synthetic network using the configuration model (inset shows a zoomed-in version of the same data). Here again it is possible to see the differences between both, due to the structure found in the empirical network. Institutions tend to collaborate less as their number of patents increase. However when the number of patents reaches around 50 per institution, a few of the latter deviate from the majority. Such institutions concentrate the information leaving the rest of the network aside.}
\label{fig:emp_vs_cm}
\end{figure*}

As the number of patents filed by an institution increases, such institutions tend to collaborate on a smaller fraction of their patents. However, this trend changes when institutions reach the number of around 50 patents, with a few prolific institutions sharing almost their entire set of patents. These institutions are those whose patents are part of the $4\%$ of all patents with the highest redundancy. This means that transfer of knowledge happens among them, but they concentrate the information, neglecting the majority of the institutions of the network. The correlation of the average collaboration ratio (ratio of shared patents) and the bipartite bottom degree (number of patents) is shown in Fig. \ref{fig:avshared_bipdegree}. 

Once more we can see the differences between the empirical network and synthetic network from the configuration model, due to the structure present in the first. In contrast to the case of the empirical network, the configuration model shows a flat average collaboration ratio, independent of the bipartite bottom degree, as a result of the rewiring process. \\

\begin{figure*}[h!]
\centering
\captionsetup{width=.82\linewidth}
\subfloat{\label{fig:simp_multi_emp} \includegraphics[scale=0.33]{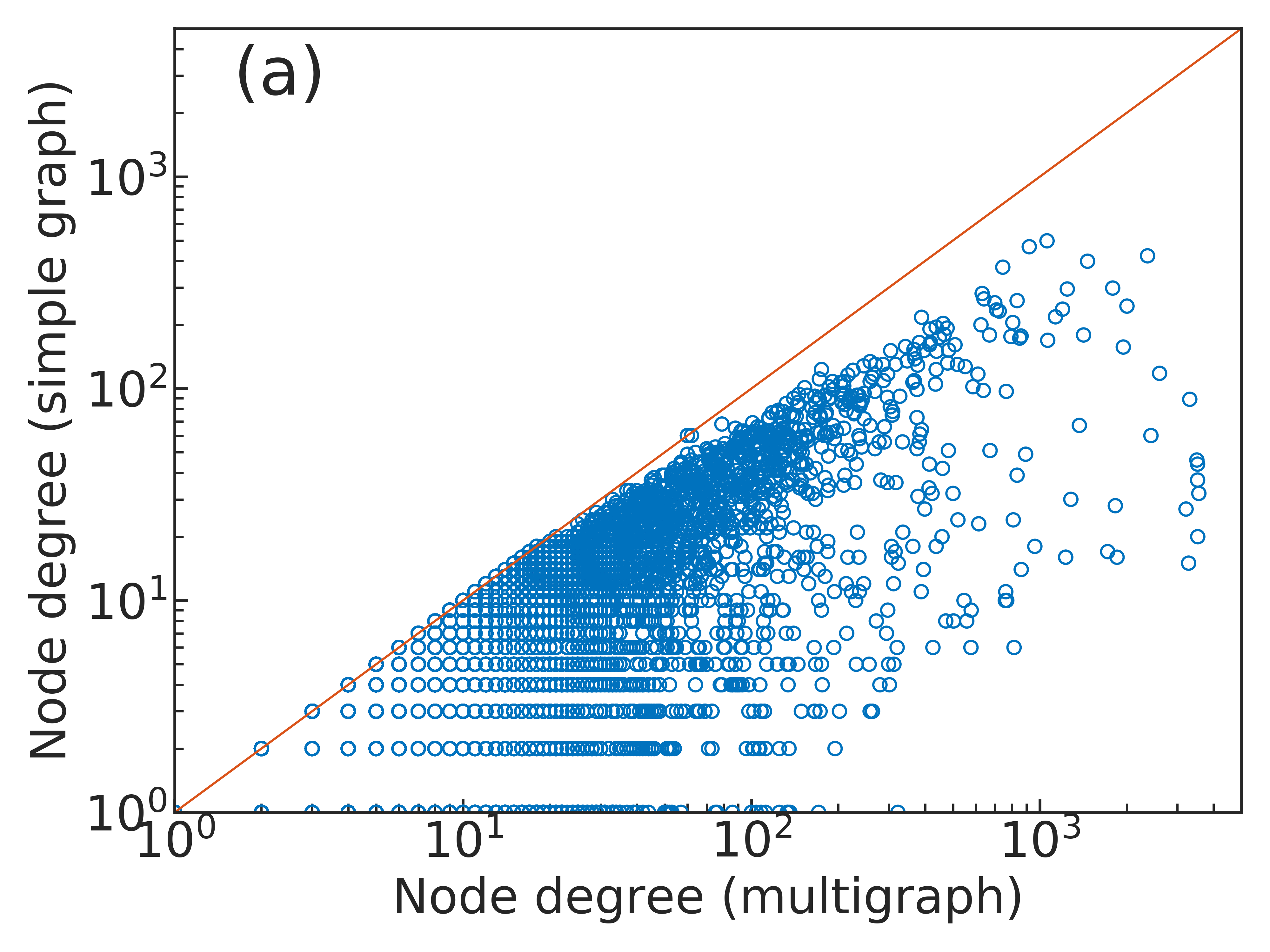}}
\subfloat{\label{fig:simp_multu_cm} \includegraphics[scale=0.33]{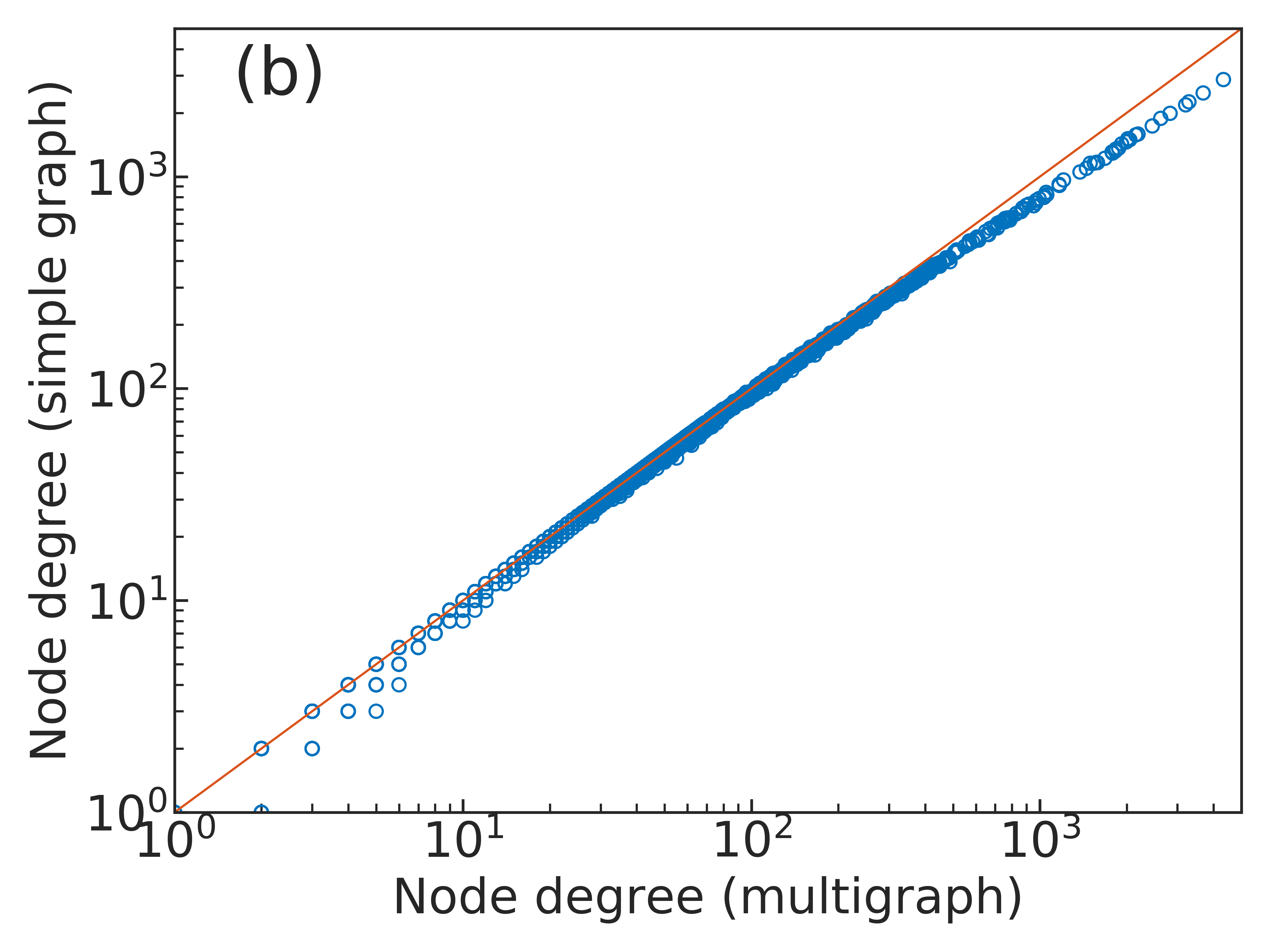}}
\caption{Correlations between nodes degrees in the simple graph projection and in the multigraph projection for (a) the empirical and (b) the synthetic network with configuration model. The red line shows the symmetry between both degrees. The more distant from the line, the more homogeneous. Due to structure breaking, the synthetic network presents bottom nodes with high heterogeneous collaborations.}
\label{fig:emp_vs_cm2}
\end{figure*}

\textbf{\textit{The Diversity of collaborators:}} is the relationship between the degree and the strength of an agent (institution) in projected networks. The maximum number of possible collaborators of a bottom node is the degree $q^{m}_{u}$ (or strength $s_{u}$) of such a node in a multigraph projection, given when Eq. (\ref{eq:qi}) is an equality, that is, $q_{u} \leq \sum_{j=1}^{k_{u}}(d_{v_j} - 1)$. On the other hand, the actual number of collaborators of $u$ is its degree $q_{u}$ in a simple graph projection. As we have already seen, this difference is due to redundancy (or four-cycle motifs) in the bipartite network. The diversity of collaborators can be expressed as 
\begin{equation}
D_{u} = \frac{q_{u}}{q^{m}_{u}} = \frac{q_{u}}{s_{u}} \,.
\end{equation}
As the number of shared patents increases and, as a consequence, the number of possible collaborators, it becomes harder for an institution to have a completely heterogeneous set of collaborators (i.e. $q_{u}=q^{m}_{u}$). In other words, prolific institutions tend to repeat collaborators over time. In Fig. \ref{fig:emp_vs_cm2} it is possible to see the diversity of collaborators via the correlation between the simple graph and multigraph degrees for every bottom node of the empirical network. The more distant points are from the line $y=x$ in red, the more homogeneous --- less diverse --- the collaborations of an agent are, meaning that the artifacts connected to this agent have bigger redundancy coefficients. A metric based in this overlapping structure found in bipartite networks, namely dispersion coefficient, was proposed in \cite{tackx2015revealing}. Although it provides the same information as the diversity measure proposed here, we believe the way the diversity is defined is simpler and more intuitive. 

In contrast to what is observed in the empirical network, most agents from the synthetic network using the configuration model are either on, or very close to, the line of symmetry. This high diversity of collaborators in the synthetic network is also due to the fact that the random rewiring process breaks the original structure present in the bipartite network.\\

\textbf{\textit{Collaborativeness:}} is a metric to capture the level of collaboration each agent has in the network. The metric is based in the degree of node $u \in U$ and on the degree of each one of its $k_{u}$ neighbours in $V$, weighted by the collaboration ratio. It is a local property given by
\begin{equation}
z_{u} = R_{u} \sum_{v_{u}=1}^{k_{u}} \ln d_{v_{u}}\,.
\end{equation}
The logarithmic function is chosen due the following reasoning:

\begin{itemize}
	\item When node $u$ is connected to a top node with degree $d_{v_{u}}=1$, there is no collaboration, and therefore such artifacts should not add any value to the collaborativeness of an agent.
	\item The function smooths the differences between low and high top nodes degrees. Let's say, for instance, that nodes $u$ and $u'$ have degree $k_{u} = k_{u'} = 1$, but the first is connected to an artifact with degree $d_{v_{u}} = 2$ and $u'$ is connected to another artifact with degree $d_{v_{u'}} = 20$. We do not want to have $z_{u'}$ being $10$ times bigger than $z_{u}$.
\end{itemize}

\begin{figure*}[h!]
\centering
\captionsetup{width=.82\linewidth}
\subfloat{\label{fig:colness_dist} \includegraphics[scale=0.33]{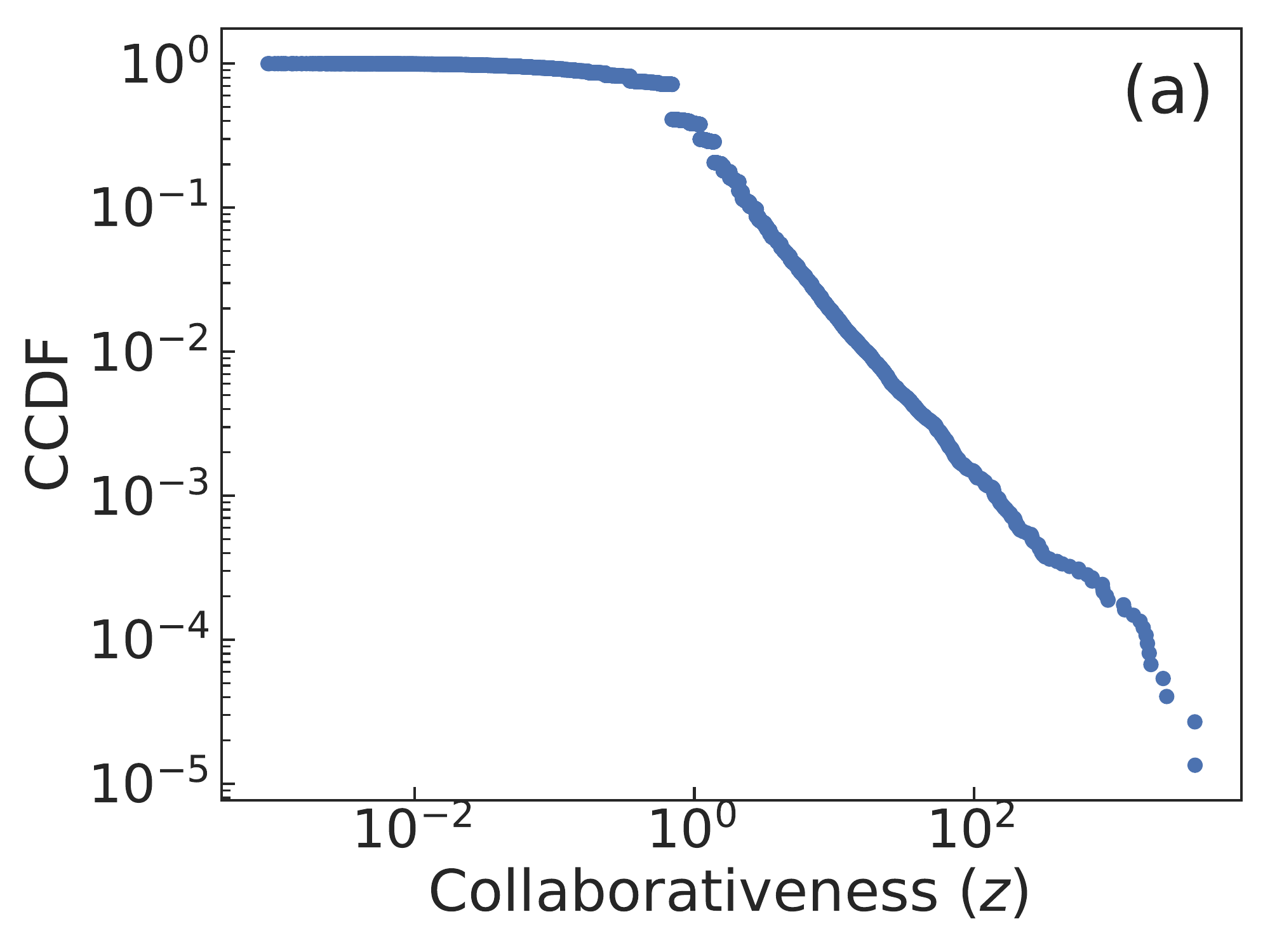}}
\subfloat{\label{fig:colness_clust} \includegraphics[scale=0.33]{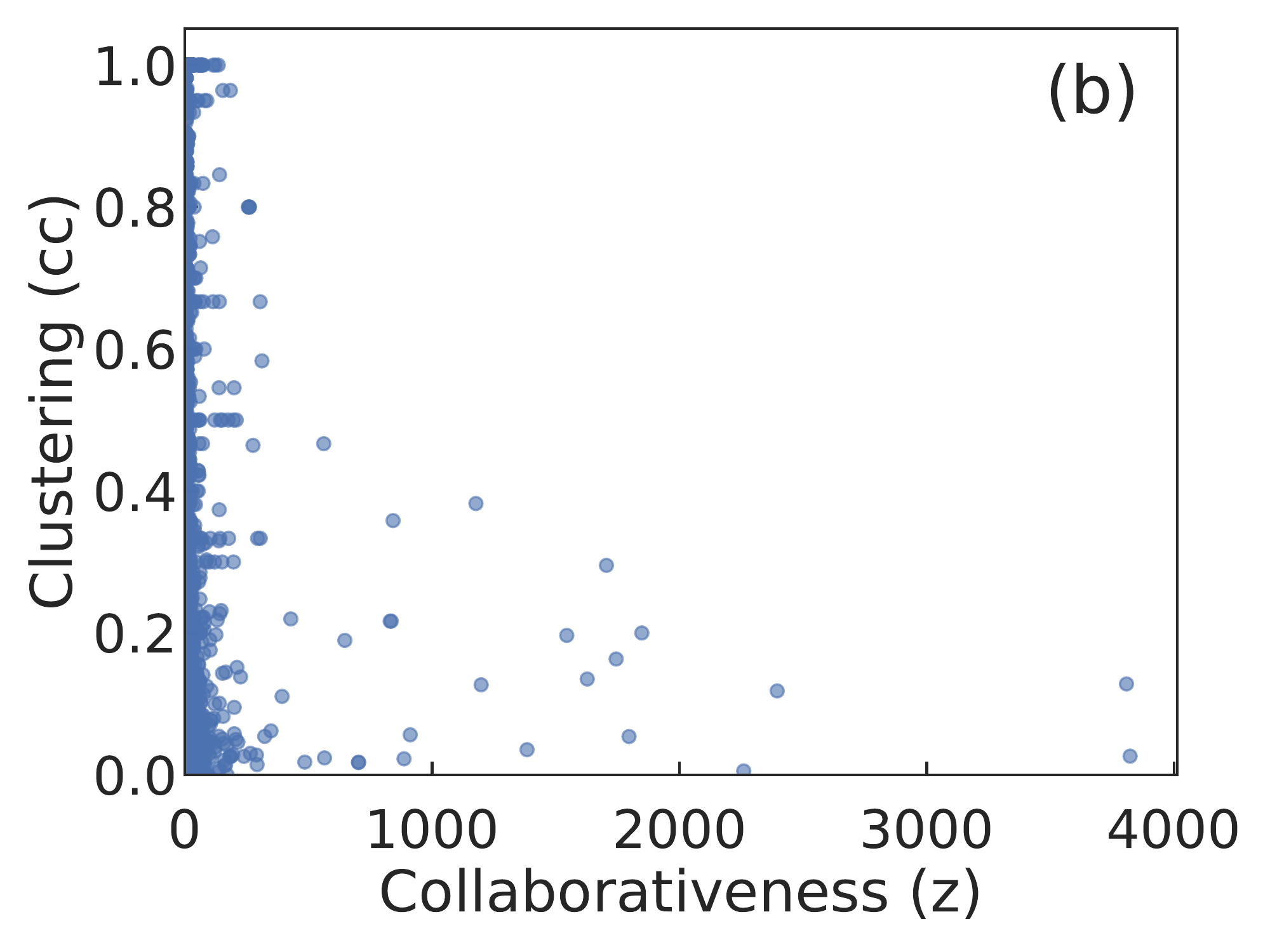}}
\caption{(a) Distribution (CCDF) for the collaborativeness values of the co-patenting network. The heavy tail of the distribution shows that high values of this metric become rapidly rare, meaning that just a small proportion of the institutions are highly collaborative. (b) Correlation between collaborativeness and clustering coefficient showing that institutions with high collaborativeness tend to have low clustering. This is not surprising as a high-degree bottom node $u$ is more likely to have high $z_u$ and low $cc_u$.}
\label{fig:colness}
\end{figure*}

The distribution of collaborativeness values is heavy-tailed (Fig. \ref{fig:colness_dist}), meaning that just a small proportion of institutions is highly collaborative. Moreover, the latter tend to have low clustering in the projected network (Fig \ref{fig:colness_clust}). Such behavior is expected as the higher the bottom node degree (number of patents) of an agent in the bipartite network, more likely it will have low clustering in the projection. High-degree bottom nodes in $B$ create more triplets in $G$ \cite{vasques2019transitivity}, increasing the denominator of Eq. (\ref{eq:global_clust}) applied to local clustering. 

Table \ref{table:top_colnes} shows the agents (institutions) with largest values of collaborativeness $z$. One can notice that such institutions are mostly different international branches of the same corporations. In addition to that, the diversity of their collaborators are frequently $0$ or very close to it. In other words, even though they collaborate on nearly the entirety of their patents (represented by the variable $R$), their collaborators are always the same. They are, indeed, collaborating almost exclusively among themselves (the international branches of the corporation). This is likely to be due to a type of ``head-office effect'' where some apparent innovation activity is in fact due to the practise of firms with multiple branches including the address of the head office on patent applications. For us, such behaviour raises a question, to be addressed in a future work, whether corporations are filing patents, with international branches, for purely bureaucratic reasons, or whether they are indeed concerned with the flow of knowledge between research teams in different countries.

\begin{table}[!ht]
\caption{Top 10 companies with largest collaborativeness in the network.}
\label{table:top_colnes}
\centering
\resizebox{\columnwidth}{!}{
\begin{tabular}{lrrrrrr}
\hline
Company			   & 
\centering Patents & 
\centering $z_{u}$     & 
\centering $R_{u}$     & 
\centering $q_{u}$     & 
\centering $q_{u}^{m}$ & 
\centering $D_{u}$ \tabularnewline
\hline
UNILEVER NV                   & 6006 & 3,822.31 & 0.96 & 28  & 5767 & 0.00 \\
UNILEVER PLC                  & 5672 & 3,807.71 & 0.98 & 13  & 5594 & 0.00 \\
PHILIPS INTELLECTUAL PROPERTY & 3430 & 2,395.28 & 1.00 & 44  & 3524 & 0.01 \\
CNRS                          & 4762 & 2,259.82 & 0.73 & 958 & 5361 & 0.18 \\
FORD MOTOR CO LTD             & 1624 & 1,847.77 & 1.00 & 20  & 3528 & 0.01 \\
FORD WERKE AG                 & 1622 & 1,795.98 & 0.98 & 46  & 3503 & 0.01 \\
SCHLUMBERGER TECH BV          & 1427 & 1,744.04 & 1.00 & 32  & 3559 & 0.01 \\
FORD FR SA                    & 1459 & 1,704.71 & 1.00 & 15  & 3280 & 0.00 \\
SERV. PETROLIERS SCHLUMBERGER & 1577 & 1,627.11 & 0.93 & 37  & 3524 & 0.01 \\
SCHLUMBERGER HOLDINGS LTD     & 1255 & 1,544.27 & 0.99 & 27  & 3213 & 0.01 \\
\hline
\end{tabular}
}
\end{table}

In contrast,  institutions with the largest number of patents in the dataset (shown in Table \ref{table:top_patents}) have a low value of collaborativeness, mostly due to low $R$ (the proportion of patents they hold, on which they collaborate is small). However, they tend to have diverse collaborators. This behavior suggests that such institutions keep their core research --- developing a large number of patents ---  in-house, while seeking out smaller institutions to collaborate with for specific projects only. 

\begin{table}[!ht]
\caption{Top 10 companies with largest number of patents in the dataset.}
\label{table:top_patents}
\centering
\resizebox{\columnwidth}{!}{
\begin{tabular}{lrrrrrr}
\hline
Company			   & 
\centering Patents & 
\centering $z_{u}$     & 
\centering $R_{u}$     & 
\centering $q_{u}$     & 
\centering $q_{u}^{m}$ & 
\centering $D_{u}$ \tabularnewline
\hline
SIEMENS AG                  & 37992 & 32.53  & 0.03 & 399 & 1464 & 0.27 \\
KON PHILIPS ELECTS NV       & 26762 & 702.56 & 0.19 & 191 & 5349 & 0.04 \\
ROBERT BOSCH GMBH           & 24863 & 14.95  & 0.03 & 176 & 794  & 0.22 \\
SAMSUNG ELECT CO LTD        & 21408 & 20.91  & 0.04 & 177 & 861  & 0.21 \\
IBM CORP                    & 21017 & 34.82  & 0.05 & 169 & 1065 & 0.16 \\
SONY CORP                   & 17833 & 43.59  & 0.06 & 237 & 1199 & 0.20 \\
CANON KK                    & 17424 & 1.85   & 0.01 & 85  & 239  & 0.36 \\
GENERAL ELECT CO            & 15893 & 0.59   & 0.01 & 85  & 132  & 0.64 \\
MATSUSHITA ELECT IND CO LTD & 14345 & 15.66  & 0.04 & 235 & 708  & 0.33 \\
TELEFON AB LM ERICSSON PUBL & 13232 & 0.74   & 0.01 & 32  & 126  & 0.25 \\
\hline
\end{tabular}
}
\end{table}

The above suggests that different strategies for collaboration are part of the reason that we observe neutral degree assortativity in the co-patenting network. We have a balance between large corporations collaborating among themselves, and the most prolific institutions collaborating with smaller ones in several different projects.  

\section{\label{sec:conclusions}Conclusion}
In this paper, we have studied the effects of bipartite networks' features in shaping one-mode projected networks' topology. We built an empirical bipartite network, using data from the European Patent Office with harmonised applicant names (OECD, HAN database, February 2016), where institutions are connected to patents they have developed. From this bipartite structure we created a projection onto the agents (bottom nodes) --- a new one-mode collaboration network linking institutions (agents) that have patented together.

We then compared the structure of the empirical network to that of synthetic networks created using random models, namely the ER $G(|U|,|L|)$ model and the configuration model, by first adapting them to bipartite networks and also by applying them straight from the one-mode projection. 
Such methods give different outcomes, and we have shown that, although the second approach (rewiring a projected network)is used more often, it is a misleading process to model projected networks. This is due to the inherent loss of information in projections. 

The first property that we looked at was the degree distributions of the networks. The top degree distribution presents a peaked Poisson-like distribution while the bottom distribution is heavy-tailed. Due to this, the projected co-patenting network structure is not degree-assortative, unlike most social networks. When we create projections using simple graph and multigraph methods, the latter has a significantly heavier tail than the former. This is due to the presence of many common neighbours between pairs of nodes in the bipartite network. That is, redundancy is high for some top nodes showing the preference of many institutions for collaborating repeatedly with others they have collaborated with in the past. To support this claim we created both projections (simple and multigraph) of a bipartite network built using the configuration model. As expected, both degree distributions are quite similar in this case. The small difference is a result of the random rearrangement of the links that create few common neighbours, due to the high degree nodes. When we rewire the links, we break the structure of the empirical network, namely the four-cycles, significantly lowering the level of redundancy of the bipartite network. 

Secondly, we turned our attention to the distribution of the clustering coefficient. Led by the fact that the configuration model of the bipartite network does not predict the simple graph degree distribution of its projection, we compared clustering levels by following different paths to model the projection. By building a one-mode network with the configuration model using the degree sequence of the empirical projected network, we preserve the degree distribution of it. However, such a model fails drastically in reproducing the clustering level of the projection. On the other hand, if we first model the bipartite network, as we did when looking at the degree distribution, and then creating the projected network, it is possible to keep some of the clustering structure, but not all of it. The additional clustering in the projected network is due to the presence of six-cycles in the bipartite network. These motifs are the representation of transitivity in the bipartite network and result in triadic closure in the projected network.

Finally, we proposed several new metrics to quantify the level and type of collaboration in the empirical co-patenting network. Using these metrics, we showed that those corporations with high relative collaborativeness tend to have a low diversity of collaborators. That is, they tend to only share patents with the same small set of collaborators --- typically with different branches of the same parent organisation. In contrast, the most prolific institutions in terms of patenting activity tend to have relatively low collaborativeness, however, their collaborations are spread over a much more diverse set of collaborators. This suggests that the most prolific institutions concentrate their R \& D activity on their core technology area while accessing complementary knowledge for specific projects by collaborating with diverse smaller institutions. These different behaviours in collaborations contribute to the neutral characteristic of the co-patenting network observed for the degree assortativity.     




\bibliographystyle{unsrt}
\bibliography{sample}
\end{document}